\documentclass[11pt]{article}
\usepackage{fullpage}
\usepackage{amsmath}
\usepackage{amssymb}
\usepackage{amsthm}
\usepackage{bbm}
\usepackage{algorithm}
\usepackage{algorithmic}
\usepackage{xcolor}
\usepackage{soul}
\usepackage{xfrac}
\usepackage{dsfont}
\usepackage{paralist}

\newcommand\fullversion{0}
\newcommand\withcomments{0}

\usepackage[colorlinks,urlcolor=blue,citecolor=blue,linkcolor=blue]{hyperref}
\usepackage{graphicx,amsfonts,amsmath,amssymb,epsfig,color,multicol,pstricks,tikz, pgf}
\usetikzlibrary{decorations.markings}
\usetikzlibrary{patterns}
\usetikzlibrary{calc, intersections}
\usepackage{pgf,tikz}
\usetikzlibrary{calc}
\usepackage{tkz-euclide}
\usepackage{pgfplots}

\newtheorem{thm}{Theorem}

\newtheorem{lem}{Lemma}[section]
\newtheorem{cor}[lem]{Corollary}

\newtheorem{clm}[lem]{Claim}

\newtheorem{dfn}[lem]{Definition}

\newtheorem{ntn}[lem]{Notation}
\newtheorem{obs}[lem]{Observation}

\DeclareRobustCommand{\hldana}[1]{{\sethlcolor{bubbles}\hl{#1}}}
\DeclareRobustCommand{\hltalya}[1]{{\sethlcolor{yellow}\hl{#1}}}
\DeclareRobustCommand{\hlwill}[1]{{\sethlcolor{pink}\hl{#1}}}
\ifnum\withcomments=1
\newcommand{\dnote}[1]{\hldana{[\textbf{D}: #1]}}
\newcommand{\dmark}[1]{\hldana{#1}}
\newcommand{\tnote}[1]{\hltalya{[\textbf{T}: #1]}}
\newcommand{\wnote}[1]{\hlwill{[\textbf{W}: #1]}}
\newcommand{\new}[1]{\textcolor{blue}{#1}}
\else
\newcommand{\dnote}[1]{}
\newcommand{\dmark}[1]{}
\newcommand{\tnote}[1]{}
\newcommand{\wnote}[1]{}
\newcommand{\new}[1]{#1}
\fi

\newcommand{\oN}{\overline{N}}
\newcommand{\tl}{t^{\low}}

\newcommand{\dt}{a}
\newcommand{\Gt}{H_k}
\newcommand{\onk}{\overline{n}_k}
\newcommand{\tnk}{\tilde{n}_k}

\newcommand{\Vt}{{V_{H_k}}}
\newcommand{\Et}{{E_{H_k}}}

\newcommand{\emphdef}[1]{\textsf{#1}}
\newcommand{\sets}{\lceil \log n\rceil }
\newcommand{\settau}{\max\left\{\frac{\alpha}{\eps}\cdot (4k\log n)^k, \left(\onk/\sqrt{n\alpha}\right)^{1/(k-1)}\right\} }

\newcommand{\setbeta}{\eps/2\lceil\log n\rceil}
\newcommand{\setGamma}{4\lceil \log^2 n\rceil/\eps}

\newcommand{\dft}[1]{\textbf{\textit{#1}}}

\newcommand{\e}{\varepsilon}
\newcommand{\eps}{\varepsilon}
\newcommand{\sO}{\tilde{O}}
\newcommand{\UE}{\mathcal{U}_E}
\newcommand{\Dd}{\calD_d}

\newcommand{\mC}{\mathcal{C}}
\newcommand{\N}{\mathbf{N}}
\newcommand{\R}{\mathbf{R}}
\newcommand{\E}{\mathbf{E}}

\DeclareMathOperator{\Aut}{Aut}

\DeclareMathOperator{\dist}{dist}
\DeclareMathOperator{\disj}{disj}
\newcommand{\EX}{\textbf{\textrm{Ex}}}
\newcommand{\abs}[1]{\left|#1\right|}
\newcommand{\norm}[1]{\left\|#1\right\|}
\newcommand{\paren}[1]{\left(#1\right)}
\newcommand{\set}[1]{\left\{#1\right\}}
\newcommand{\sqb}[1]{\left[#1\right]}
\newcommand{\sucht}{\,\middle|\,}
\renewcommand{\th}{^{\textrm{th}}}

\newcommand{\eqdef}{\stackrel{\text{def}}{=}}

\newcommand{\se}{\hyperref[se]{\textup{\color{black}{\sf Sample-Edge}}}}
\newcommand{\SBE}{\hyperref[SBE]{\textup{\color{black}{\sf Sample-Basic-Edge}}}}

\newcommand{\sclq}{\hyperref[sclq]{\textup{\color{black}{\sf Sample-a-Clique}}}}
\newcommand{\slz}{\hyperref[slz]{\textup{\color{black}{\sf Sample-$E(L_0)$}}}}
\newcommand{\snbr}{\hyperref[snbr]{\textup{\color{black}{\sf Sample-a-Neighbor}}}}
\newcommand{\ISLZ}{\hyperref[ISLZ]{\textup{\color{black}{\sf Define-$L_0$}}}}

\DeclareMathOperator{\poly}{poly}
\DeclareMathOperator{\tvd}{dist_{TV}}

\renewcommand{\deg}{{\sf deg}}
\DeclareMathOperator{\nbr}{{\sf nbr}}
\DeclareMathOperator{\pair}{{\sf pair}}
\DeclareMathOperator{\answer}{{\sf ans}}

\newcommand{\mU}{\mathcal{U}}
\newcommand{\mA}{\mathcal{A}}

\newcommand{\mV}{\mathcal{V}}

\newcommand{\mhat}{\widehat{m}}

\newcommand{\HP}{P}

\newcommand{\mL}{\mathcal{L}}
\newcommand{\emh}{E_{\mH}}
\newcommand{\eml}{E_{\mL}}
\newcommand{\tthres}{(18\ot)^{2/3}/\eps^{1/3}}
\newcommand{\smh}{\sqrt{\eps m/2}}
\newcommand{\mT}{\mathcal{T}}
\newcommand{\mH}{\mathcal{H}}
\newcommand{\tH}{\mT_{\mH}}
\newcommand{\mtl}{\mT_{\mL}}

\newcommand{\dm}{d^{\mL}}
\newcommand{\dpl}{d^{\mH}}
\newcommand{\mTh}{T_{\mathcal{H}}}
\newcommand{\mTl}{T_{\mathcal{L}}}
\newcommand{\mls}{m_{\ell}(S)}
\newcommand{\tls}{t_{\ell}(S)}
\newcommand{\els}{E_{\mL}(S)}
\renewcommand{\tt}{\tilde{t}}

\newcommand{\FAIL}{{\sc FAIL}}

\newcommand{\calE}{\mathcal{E}}
\DeclareMathOperator{\cost}{cost}
\newcommand{\ds}{\Delta_*}
\newcommand{\qs}{q_*}
\newcommand{\vs}{v_*}
\newcommand{\ls}{\ell_*}
\newcommand{\tk}{T^{kn}}
\newcommand{\Vk}{V^{kn}}
\newcommand{\ts}{T_{\sigma}}

\newcommand{\cA}{\mathcal{A}}

\newcommand{\calP}{\mathcal{P}}

\newcommand{\loglog}{{\rm loglog}}

\pgfdeclarelayer{background}
\pgfsetlayers{background,main}

\tikzset{dotted pattern/.style args={#1}{
		postaction=decorate,
		decoration={
			markings,
			mark=between positions 0.25 and 0.75 step 0.25 with {
				\fill[radius=#1] (0,0) circle;
			}
		}
	},
	dotted pattern/.default={1pt},
}

\tikzstyle{none}=[]

\tikzstyle{small}=[draw,circle,inner sep=0.1cm, fill=black]
\tikzstyle{blue}=[draw,circle,inner sep=.1cm, fill=blue]
\tikzstyle{green}=[draw,circle,inner sep=0.1cm, fill=violet]
\tikzstyle{red node}=[draw,circle,inner sep=0.1cm, fill=red]
\tikzstyle{gray}=[draw,circle,inner sep=0.1cm, fill=gray]
\tikzstyle{light}=[black]
\tikzstyle{new}=[black]

\newcommand{\drawSwitch}{
\begin{tikzpicture}
\begin{scope}[scale=.75]
		\node[] (A) at (-2.25, 6.75) {$A$};
		\node[] (C) at (2, 6.75) {$C$};
		\node[] (A) at (6.25, 6.75) {$A$};
		\node[] (C) at (10.25, 6.75) {$C$};
		\node [style=none] (93) at (0, -4) {$\overline{w}$};
		\node [style=none] (93) at (6.5, -4) {${w}$};
		\node [style=blue] (0) at (-3.25, -1.5) {};
		\node [style=green, label=above:{$y_0$}] (1) at (-2.25, 0.75) {};
		\node [style=green, label=above:{$y_1$}] (2) at (-1, -1.25) {};
		\node [style=blue] (3) at (-1, -2.75) {};
		\node [style=blue] (4) at (-3, -3) {};
		\node [style=red node, label=above:{$u_0$}] (8) at (-1.5, 0.5) {};
		\node [style=red node, label=above:{$u_1$}] (9) at (-3.5, -0.25) {};
		\node [style=gray] (10) at (2, 3.5) {};
		\node [style=gray] (11) at (2, 3) {};
		\node [style=gray] (12) at (2, 2) {};
		\node [style=gray] (13) at (2, 0.75) {};
		\node [style=gray] (14) at (2, 0.25) {};
		\node [style=gray] (15) at (2, -0.5) {};
		\node [style=gray] (16) at (2, -0.5) {};
		\node [style=blue] (17) at (5.25, -1.5) {};
		\node [style=green, label=above:{$y_0$}] (18) at (6.25, 0.75) {};
		\node [style=green, label=above:{$y_1$}] (19) at (7, -1.25) {};
		\node [style=blue] (20) at (7.5, -2.75) {};
		\node [style=blue] (21) at (5.5, -3) {};
		\node [style=red node, label=above:{$u_0$}] (25) at (7.5, 0.5) {};
		\node [style=red node, label=above:{$u_1$}] (26) at (5.75, 0) {};
		\node [style=gray] (27) at (10.5, 3.5) {};
		\node [style=gray] (28) at (10.5, 3) {};
		\node [style=gray] (29) at (10.5, 2) {};
		\node [style=gray] (30) at (10.5, 0.75) {};
		\node [style=gray] (31) at (10.5, 0.25) {};
		\node [style=gray] (32) at (10.5, -0.5) {};
		\node [style=gray] (33) at (10.5, -0.5) {};
		\node [style=blue] (34) at (-2.25, 5.5) {};
		\node [style=gray] (35) at (2, 6.25) {};
		\node [style=gray] (36) at (2, 5.75) {};
		\node [style=gray] (37) at (2, 4.75) {};
		\node [style=blue] (40) at (6.25, 5.75) {};
		\node [style=gray] (41) at (10.5, 6.25) {};
		\node [style=gray] (42) at (10.5, 5.75) {};
		\node [style=gray] (43) at (10.5, 4.75) {};
		\node [style=blue] (44) at (-2.25, 4.5) {};
		\node [style=none] (45) at (-1, 4.75) {};
		\node [style=none] (46) at (-1, 4.5) {};
		\node [style=none] (47) at (-1, 4.25) {};
		\node [style=blue] (48) at (6.25, 5) {};
		\node [style=none] (49) at (7.5, 5.25) {};
		\node [style=none] (50) at (7.5, 5) {};
		\node [style=none] (51) at (7.5, 4.75) {};
		\node [style=blue] (52) at (6.25, 3.75) {};
		\node [style=none] (53) at (7.5, 4) {};
		\node [style=none] (54) at (7.5, 3.75) {};
		\node [style=none] (55) at (7.5, 3.5) {};
		\node [style=blue] (56) at (-2.25, 3.5) {};
		\node [style=none] (57) at (-1, 3.75) {};
		\node [style=none] (58) at (-1, 3.5) {};
		\node [style=none] (59) at (-1, 3.25) {};
		\node [style=none] (60) at (4.75, 4.25) {};
		\node [style=none] (87) at (10.75, 4.25) {};
	\path[dotted pattern] (-2.25,3) to (-2.25,1.7);
		\path[dotted pattern] (6.25,3) to (6.25,1.7);
		\path[dotted pattern] (11) to (12);
		\path[dotted pattern] (36) to (37);		
		\path[dotted pattern] (14) to (16);
		\path[dotted pattern] (31) to (32);	
		\draw [style=light] (3) to (0);
		\draw [style=light] (0) to (4);
		\draw [style=light] (4) to (3);
		\draw [style=light] (3) to (2);
		\draw [style=light] (2) to (0);
		\draw [style=light] (0) to (1);
		\draw [style=light] (1) to (2);
		\draw [style=light] (2) to (4);
		\draw [style=light] (1) to (3);
		\draw [style=light] (1) to (4);
		\draw [style=light] (16) to (9);
		\draw [style=light] (9) to (14);
		\draw [style=light] (9) to (13);
		\draw [style=light] (8) to (12);
		\draw [style=light] (8) to (11);
		\draw [style=light] (8) to (10);
		\draw [style=light] (20) to (17);
		\draw [style=light] (17) to (21);
		\draw [style=light] (21) to (20);
		\draw (18) to (27);
		\draw (18) to (28);
		\draw (18) to (29);
		\draw (19) to (30);
		\draw (19) to (31);
		\draw (19) to (33);
		\draw (26) to (17);
		\draw (26) to (21);
		\draw (26) to (20);
		\draw (25) to (26);
		\draw (25) to (17);
		\draw (25) to (21);
		\draw (25) to (20);
		\draw [style=light] (34) to (37);
		\draw [style=light] (34) to (36);
		\draw [style=light] (34) to (35);
		\draw [style=light] (40) to (43);
		\draw [style=light] (40) to (42);
		\draw [style=light] (40) to (41);
		\draw [style=light] (44) to (47.center);
		\draw [style=light] (44) to (46.center);
		\draw [style=light] (44) to (45.center);
		\draw [style=light] (48) to (51.center);
		\draw [style=light] (48) to (50.center);
		\draw [style=light] (48) to (49.center);
		\draw [style=light] (52) to (55.center);
		\draw [style=light] (52) to (54.center);
		\draw [style=light] (52) to (53.center);
		\draw [style=light] (56) to (59.center);
		\draw [style=light] (56) to (58.center);
		\draw [style=light] (56) to (57.center);

\end{scope}
\end{tikzpicture}
}

\newcommand{\drawLB}{
\begin{tikzpicture}
\begin{scope}[scale=.75]

\node (p1) at (-7, -4) {};
\node (p2) at (-11, -5) {};
\node (p3) at (-7, 1) {};
\node (p4) at (-10, 2) {};
\draw plot [smooth cycle,tension=1] coordinates {(p1) (p2) (p4) (p3)};
\node (p5) at (-8.8, 3.5) {$D$};
\node [] (p6) at (-8.5, 0) {$n'$  vertices,};
\node [] (p6) at (-8.5, -.7) {$\Theta(n_3)$ cliques,};
\node [] (p6) at (-8.5, -1.4) {arboricity$\leq \alpha$ };

\node [style=red node, label=left:{$b_1$}] (0) at (-3, 2.5) {};
\node [style=red node, label=left:{$b_i$}] (3) at (-3, -1) {};
\node [style=red node, label=left:{$b_n$}] (4) at (-3, -4) {};
\node [style=blue, label=above:{$a_1$}] (5) at (2, 1.25) {};
\node [style=blue] (6) at (2, -2) {};
\node [style=blue] (7) at (2, -3.75) {};
\node [style=gray] (8) at (5.5, 2.25) {};
\node [style=gray] (9) at (5.5, 1.75) {};
\node [style=gray] (10) at (5.5, 0.75) {};
\node [style=gray, label=right:{$c^{a_j}_{1}$}] (11) at (5.5, -1.25) {};
\node [style=gray] (12) at (5.5, -1.75) {};
\node [style=gray, label=right:{$c^{a_j}_{3}$}] (13) at (5.5, -2.75) {};
\node [style=none] (14) at (-3, 4.25) {};
\node [style=none] (15) at (-3, 3.5) {$B$};
\node [style=none] (16) at (2, 3.5) {$A$};
\node [style=none] (17) at (5.5, 3.5) {$C$};
\node [style=none] (20) at (0.75, 2) {};
\node [style=none] (21) at (0.75, 0.5) {};
\node [style=none] (22) at (-1.75, -0.25) {};
\node [style=none] (23) at (-1.75, -1.75) {};
\node [style=none] (24) at (-2.175, -0.5) {};
\node [style=none] (25) at (-2.175, -1.5) {};
\node [style=none] (27) at (1.1, 0.75) {};
\node [style=none] (28) at (1.1, 1.75) {};
\node [style=green] (29) at (1, -5) {};
\node [style=blue] (30) at (2, -6.25) {};
\node [style=blue] (31) at (3, -5) {};
\node [style=green] (32) at (1, -5) {};
\node [style=none] (33) at (-0.25, -4.25) {};
\node [style=none] (34) at (-0.25, -5.75) {};
\node [style=none] (35) at (0.175, -5.5) {};
\node [style=none] (36) at (0.175, -4.5) {};
\node [style=none] (37) at (3, -3.25) {S};
\node [style=blue] (38) at (1, -5) {};
\node [style=blue] (39) at (2, -0.25) {};
\node [style=blue] (40) at (2, -1.25) {};
\node [style=none] (41) at (1.25, 1.25) {};
\node [style=none] (42) at (1.25, 1.25) {};
\node [style=none] (43) at (1.375, 1.25) {$\ell$};
\node [style=none] (44) at (-2.325, -1) {$\alpha$};
\node [style=none] (47) at (0.25, -5) {$\ell$};
\node [style=none] (48) at (-0.25, -1.75) {};
\node [style=none] (49) at (7.25, 2.25) {};
\node [style=none] (50) at (6.75, 2.25) {};
\node [style=none] (52) at (-1, 3.25) {};

\node [style=none] (531) at (3.6, 1.7) {};
\node [style=none] (541) at (3.575, 1.025) {};
\node [style=none] (561) at (3.15, 1.25) {\small{$k-1$}};

		\node [style=none, label=below:{$a_j$}] (53) at (2, -2) {};
\node [style=none] (54) at (0.75, -1.25) {};
\node [style=none] (55) at (0.75, -2.75) {};
\node [style=none] (56) at (1.1, -2.5) {};
\node [style=none] (57) at (1.1, -1.5) {};
\node [style=none] (58) at (1.25, -2) {};
\node [style=none] (59) at (1.25, -2) {};
\node [style=none] (60) at (1.375, -2) {$\ell$};

\draw [style=light, in=195, out=15] (5) to (8);
\draw [style=light] (5) to (9);
\draw [style=light] (5) to (10);
\path [dotted pattern] (10)+(0,-0.5) to (11)+(0,0.1);
\draw [style=light] (6) to (11);
\draw [style=light] (6) to (12);
\draw [style=light] (6) to (13);
\draw (22.center) to (3);
\draw (3) to (23.center);
\draw (21.center) to (5);
\draw (5) to (20.center);
\draw [bend right=45] (25.center) to (24.center);
\draw [in=135, out=-135] (28.center) to (27.center);
\draw [style=new, in=270, out=90] (30) to (7);
\draw [style=new] (7) to (29);
\draw [style=new] (29) to (30);
\draw [style=new] (30) to (31);
\draw [style=new] (31) to (7);
\draw [style=new, in=360, out=180] (31) to (29);
\draw (34.center) to (32);
\draw (32) to (33.center);
\draw [bend right=45] (36.center) to (35.center);
		\path [dotted pattern] (12) to (13);
\path [dotted pattern] (9) to (10);
\path [dotted pattern] (0)+(0,-.5) to (3)+(0,1.5);
\path [dotted pattern] (3)+(0,-.5) to (4)+(0,1.5);
		\draw [bend left, looseness=1] (531.center) to (541.center);

		\draw (55.center) to (53);
\draw (53) to (54.center);
\draw [in=135, out=-135] (57.center) to (56.center);

\end{scope}
\end{tikzpicture}
}

\definecolor{bubbles}{rgb}{0.91, 1.0, 1.0}
\definecolor{amethyst}{rgb}{0.6, 0.4, 0.8}

\title{Almost Optimal Bounds for Sublinear-Time Sampling of $k$-Cliques: \\ Sampling Cliques is Harder Than Counting}
\author{
  Talya Eden\thanks{CSAIL, MIT, Cambridge, MA, USA}
  \and
  Dana Ron\thanks{School of Electrical Engineering, Tel Aviv University, Tel Aviv, Israel}
  \and
  Will Rosenbaum\thanks{Amherst College, Amherst, MA, USA}
}

\begin{document}

\begin{titlepage}

\maketitle

\begin{abstract}

\tnote{REMOVE COMMENTS}
Motivated by the need to analyze massive datasets efficiently, recent theoretical work has examined the problems of counting and sampling small subgraphs from graphs in sublinear time. In this work, we consider the problem of sampling a $k$-clique in a graph from an almost uniform distribution.
Specifically the algorithm should output each $k$-clique with probability $(1\pm \epsilon)/n_k$, where $n_k$ denotes the number of $k$-cliques in the graph and $\epsilon$ is a given approximation parameter. To this end it may perform the following types of queries on vertices:
degree queries, neighbor queries, and  pair queries.

We prove that the query complexity of this problem is
\[
\Theta^*\left(\max\left\{ \left(\frac{(n\alpha)^{k/2}}{ n_k}\right)^{\frac{1}{k-1}} ,\; \min\left\{n\alpha,\frac{n\alpha^{k-1}}{n_k} \right\}\right\}\right).
\]
where $n$ is the number of vertices in the graph, $\alpha$ is its arboricity,
and $\Theta^*$ suppresses the dependence on $(\log n/\epsilon)^{O(k)}$.
Interestingly, this establishes a separation between approximate counting and approximate uniform sampling in the sublinear regime. For example, if $k=3$, $\alpha = O(1)$, and $n_3$ (the number of triangles) is $\Theta(n)$, then we get a lower bound of $\Omega(n^{1/4})$ (for constant $\epsilon$), while under these conditions, a $(1\pm \epsilon)$-approximation of $n_3$ can be obtained by performing $\poly(\log(n/\epsilon))$ queries (Eden, Ron and Seshadhri, SODA20).

Our lower bound follows from a construction of a family of graphs with arboricity $\alpha$ such that in each graph there are $n_k$ cliques (of size $k$), where one of these cliques is ``hidden'' and hence hard to sample. Our upper bound is based on defining a special auxiliary graph $H_k$, such that sampling edges almost uniformly in $H_k$ translates to sampling $k$-cliques almost uniformly in the original graph $G$. We then build on a known edge-sampling algorithm (Eden, Ron and Rosenbaum, ICALP19) to sample edges in $H_k$, where the challenge is simulate queries to $H_k$ while being given access only to $G$.

\end{abstract}

\thispagestyle{empty}
\end{titlepage}


\setcounter{page}{1}
\newpage

\section{Introduction}

Counting and sampling are fundamental computational tasks in randomized algorithms, statistics, data science, and many other disciplines. Given a family $\mathcal{F}$ of combinatorial objects---for example, $k$-cliques in a given graph, or satisfying assignments of a Boolean formula---the \emph{approximate counting problem} asks 
to compute a number that is a $(1 \pm \eps)$-multiplicative estimate of $N = \abs{\mathcal{F}}$. The \emph{almost uniform sampling problem} is to produce a sample from $\mathcal{F}$ such that each $x \in \mathcal{F}$ is chosen with probability in the range $(1 \pm \eps) / N$.\footnote{We refer to this notion of ``almost uniform'' sample as \emph{pointwise} almost uniformity. Pointwise uniformity is the notion used, for example, in~\cite{jerrum1986random}, and is a strictly stronger requirement than approximate uniformity with respect to $L_2$ or total variation distance (TVD).} In a seminal work Jerrum, Valiant, and Vazirani~\cite{jerrum1986random} showed that for a large family of combinatorial problems---self-reducible problems~\cite{schnorr1981self-transformable}---approximate counting and almost uniform sampling are equivalent under polynomial-time reductions.
In~\cite{DLM20}, Dell, Lapinskas, and Meeks  proved related results in the fine-grained setting.\footnote{Namely, they prove “black box” results for turning algorithms which decide whether or not a witness exists into algorithms to approximately count the number of witnesses, or to sample from the set of witnesses approximately uniformly, with essentially the same running time.}

Recently, Fichtenberger, Gao, and Peng~\cite{Peng} asked if results analogous to~\cite{jerrum1986random} hold for sublinear-time algorithms:
\begin{quote}
  \emph{In the sublinear-time regime, is almost uniform sampling `computationally comparable' to approximate counting?}
\end{quote}
In~\cite{Peng}, the authors provide evidence for a positive answer to this question. Building upon previous results of Assadi, Kapralov, and Khanna~\cite{AKK19}, Fichtenberger et al.\ describe algorithms for approximately counting and sampling arbitrary subgraphs of a graph $G$. In the case of counting and sampling $k$-cliques, both algorithms have expected run-time $O^*(m^{k/2} / n_k)$,\footnote{We use $\Theta^*$ to suppress a dependence on functions $g(\log n,k,1/\eps)$, which are at most $(\log n/\eps)^{O(k)}$.} where $m$ and $n_k$ denote the number of edges and $k$-cliques in $G$, respectively.\footnote{We note that the upper bound for approximate counting of $k$-cliques was already known due to~\cite{ERS20-clqs}.} These upper bounds are essentially optimal by a nearly-matching lower bound due to Eden and Rosenbaum~\cite{ER18_LB}.




The algorithms of~\cite{Peng} and~\cite{AKK19}, however, require a non-standard ``augmented'' query model that allows the algorithm to sample a uniformly random \emph{edge} from the graph as an atomic unit-cost query. This model is strictly stronger than the well-studied \emph{general graph model}, which allows for only degree, neighbor, and pair queries. Indeed, Eden and Rosenbaum~\cite{ER18_LB} prove a lower bound of $\Omega\left(n / n_k^{1/k} + m^{k/2} / n_k \right)$ for the query complexity of approximately counting $k$-cliques in the general graph model, which is strictly greater than the upper bounds of~\cite{AKK19} and~\cite{Peng} for some range of parameters. Thus, it is not clear that counting and sampling should have the same query complexities in the general graph model.

Another subtlety not addressed in~\cite{ERS20-clqs,AKK19,Peng} is that the complexity of approximate counting can vary dramatically for restricted families of input graphs. The algorithms of~\cite{ERS20-clqs,AKK19,Peng} are only optimal when considering the worst-case over all possible inputs. However for a rich family of graphs---namely the family of graphs with bounded arboricity\footnote{The arboricity of a graph $G$, denoted $\alpha(G)$, is the minimal number of forests required to cover its edge set. It is well known, that up to a factor of 2, it is equivalent to the average degree of the densest subgraph in $G$.}---the lower bounds can be circumvented:
In~\cite{ERS20_soda}, Eden, Ron and Seshadhri prove that when given a bound $\alpha$ on the arboricity of the input graph $G$, the value of $n_k$ can be approximated in time\footnote{Recall that for every graph $G$ with arboricity at most $\alpha$, $m$ is always upper bounded by $\alpha$.}
\begin{equation}
  \label{eqn:ers20}
O^*\left(
\min\left\{\frac{n}{n_k^{1/k}}, \frac{n\alpha^{k-1}}{n_k} \right\}+
\frac{m \alpha^{k-2}}{n_k}
\right)
=O^*\left(\frac{n\alpha^{k-1}}{n_k}\right).
\end{equation}
For certain ranges of the parameters, this upper bound is exponentially smaller than the worst-case lower for general graphs. A natural question is whether a similar result can be established for sampling $k$-cliques almost uniformly in bounded arboricity graphs.



In~\cite{ERR19}, Eden, Rosenbaum and Ron provide a positive answer to this question in the case of sampling edges (i.e., $2$-cliques) in a graph. Specifically, they prove that the query complexity of uniformly edge is $\Theta^*\left(n \alpha / m\right)=\Theta^*\left(n \alpha / n_2\right)$. This complexity matches the complexity of~(\ref{eqn:ers20}) for $k = 2$, thereby providing a positive answer to Fichtenberger et al.'s question in the case of counting and sampling edges. Moreover, the tight correspondence between the complexities of counting and sampling edges holds even when parameterized by the graph's arboricity.

In the current work, we show that, surprisingly, the tight correspondence for the complexity of sampling and counting $k$-cliques does not generalize to $k \geq 3$. Specifically, we prove the following theorem.

\begin{thm}\label{thm:lb}
Any almost uniform $k$-cliques sampling algorithm for graphs with arboricity at most $\alpha$ requires
\[
\Omega\left(\max\left\{ \left(\frac{(n\alpha)^{k/2}}{k^k\cdot n_k}\right)^{\frac{1}{k-1}} ,\; \min\left\{n\alpha,\frac{n\alpha^{k-1}}{n_k} \right\}\right\}\right)
\]
queries.
\end{thm}
The second term in the lower bound follows directly from the lower bound of~\cite{ERS20_soda} for the approximate counting variant of the problem. The first term, however, might be significantly larger. For example, for the case of triangles ($k=3$), $\alpha=O(1)$, and $n_3=\Theta(n)$, the first term translates to a a lower bound of $\Omega(n^{1/4})$ for approximately uniform sampling. This is in stark contrast to the counting variant which has complexity $O^*(1)$, implying an exponential gap between the two tasks for certain regimes of parameters.

While this lower bound on the complexity might seem unnatural at first glance, we also prove an almost-matching upper bound, thus resolving the complexity of the problem up to $(\log n/\eps)^{O(k)}$ factors.

\begin{thm}\label{thm:ub_informal}
	There exists an almost uniform sampling algorithm for $k$-cliques in graphs with arboricity at most $\alpha$.
	  Given a constant factor estimate of $n_k$,  the query complexity of the algorithm is
	\[
	O^*\left(\max\left\{ \left(\frac{(n\alpha)^{k/2}}{ n_k}\right)^{\frac{1}{k-1}} ,\; \min\left\{n\alpha,\frac{n\alpha^{k-1}}{n_k} \right\}\right\}\right).
	\]
\end{thm}
If the algorithm is not provided with an estimate of $n_k$, then an estimate of $n_k$ can be obtained by applying the algorithm of~\cite{ERS20_soda} whose expected query complexity~(\ref{eqn:ers20}) is dominated by the runtime of Theorem~\ref{thm:ub_informal}.

\paragraph{Remarks on almost uniformity.} In the results listed above, we measure ``almost uniformity'' with respect to pointwise distance between distributions. That is, we require that \emph{every} $k$-clique is sampled with probability $(1 \pm \eps) / n_k$. One could also consider the (strictly weaker) requirement that the distribution of sampled $k$-cliques is close to uniform with respect to total variation distance (TVD). That is, if $p_C$ is the probability that the algorithm returns a clique $C$, then $\sum_C \abs{p_c - 1/n_k} \leq \eps$. Interestingly, the complexity of sampling $k$-cliques almost uniformly with respect to TVD is different from the bounds prescribed in Theorems~\ref{thm:lb} and~\ref{thm:ub_informal}. The algorithm of Eden, Ron, and Seshadhri~\cite{ERS20_soda} can be adapted to sample a $k$-clique almost uniformly with respect to TVD using the same number of queries
(stated in Equation~\eqref{eqn:ers20})
as their approximate counting algorithm.
In particular, combining this observation with Theorem~\ref{thm:lb} gives an exponential separation between the complexities of sampling $k$-cliques (1) pointwise almost uniformly, and (2) almost uniformly with respect to TVD, when $k \geq 3$. In contrast, for the case of edges ($k = 2$), TVD and pointwise almost uniform sampling can both have sample complexity $O^*(1)$ in bounded arboricity graphs~\cite{ERR19}.

Sampling almost uniformly with respect to TVD may be sufficient in many contexts.
However, there are scenarios in which we  the stronger notion of pointwise almost uniform sampling is crucial, since we cannot allow to ``give up'' on sampling a small fraction of the domain elements, as might be the case when sampling with respect to TVD.
See~\cite{ER18_edges} for further discussion.

\ifnum\fullversion=1

\section{Related Work}\label{sec:related-work}

We note that some of the works were mentioned before, but we repeat them here for the sake of completeness.
In recent years there has been an increasing interest in the questions of subgraph approximate counting and uniform sampling in sublinear-time.
The works differ by the query model, graph class of $G$ and the
subgraph $H$ at question.

\textbf{The general graph query model.}
The first works on estimating the number of subgraph counts were by Feige~\cite{feige2006sums} and Goldreich and Ron~\cite{GR08}, who
presented algorithms for approximately counting the number of $k$-cliques in a graph for $k=2$ (edges).\footnote{Feige considered a model that only allows for degree queries, and presented a factor $2$ approximation algorithm, and also proved that with no additional queries this approximation factor cannot be improved in sublinear time. Goldreich and Ron then considered this question allowing also for neighbor queries. In this model the proved an $(1\pm\eps)$-factor approximation algorithm with the same complexity as the previous one (as well as a matching lower bound).} Later, Gonen, Ron and Shavitt~\cite{GRS11}
 gave essentially optimal bounds for the problem of approximately counting the number of stars in a graph.
 In~\cite{ELRS-focs, ERS20-clqs}  Eden, Levi, Ron and Seshadhri and Eden, Ron and Seshadhri presented essentially optimal bounds for the problems of approximately counting triangles and $k$-cliques.

 \textbf{Augmented model.}
 In~\cite{Aliak}, Aliakbarpour, Biswas, Gouleakis, Peebles, and Rubinfeld and Yodpinyanee suggested a model that also allows for uniform edge samples. In that model they presented improved bounds for the approximate star counting problem.
 In that model, Assadi, Kapralov and Khanna~\cite{AKK19} considered the problem of approximate counting of arbitrary subgraphs $H$. The expected query complexity of their algorithm is $\frac{m^{\rho(H)}}{n_H}$, where $\rho(H)$ is the fractional edge cover of $H$\footnote{The fractional edge cover of a graph $H=(V_H, E_H)$ is a mapping $\psi:E_H \rightarrow [0,1]$ such that for each vertex $a \in V_H$, $\sum_{e \in E_H, a \in e} \psi(e) \geq 1.$ The fractional edge-cover number $\rho(H)$ of $H$ is the minimum value	of $\sum_{e \in E_H} \psi(e)$ among all fractional edge covers $\psi$.}, and $n_H$ is the number of copies of $H$ in $G$. Their result is optimal for the case of $k$-clique and odd-cycle counting.

 In~\cite{ER18_LB}, Eden and Rosenbaum presented a framework for proving subgraph counting lower bounds using reduction from communication complexity, which allowed them to reprove the lower bounds for all of the variants listed above.

\textbf{Set query model.}
In~\cite{beame2017edge},  Beame, Har{-}Peled, Ramamoorthy and
Sinha suggested two new models that allow what they refer to as \emph{independent set} (IS) and \emph{bipartite independent set} (BIS) queries. They considered the problem of estimating the number of edges and gave $O^*(n^{2/3})$ and $O^*(1)$ algorithms for this problem using IS and BIS queries, respectively. The first result was later improved by Chen, Levi and Waingarten~\cite{Chen-IS-edges} who settled the complexity of the problem to $\Theta^*(n/\sqrt m)$.
In~\cite{bhattacharya2019triangle}, Bhattacharya, Bishnu, Ghosh, and Mishra later have generalized the BIS model to tripartite set queries, where they considered the problem of triangle counting.

\textbf{Uniform sampling.}
In~\cite{ER18_edges}, Eden and Rosenbaum initiated the study of sampling subgraphs (almost) uniformly at random. They considered the general graph query model, and  presented upper and matching lower bounds for the problem of sampling edges almost uniformly. Their algorithm matches the complexity of the counting variant of the problem. Their  algorithm's dependency on $\eps$ was later improved by T\v{e}tek~\cite{Tetek}, so that the new algorithm allos sampling from the \emph{exact} uniform distribution.
In~\cite{Peng}, Fichtenberger, Gao and Peng proved that in the augmented edge model, \emph{exact} uniform sampling of arbitrary subgraphs can be performed in $O\left(\frac{m^{\rho(H)}}{n_H}\right)$ time. This matches the upper bound of ~\cite{AKK19} for the counting variant.

\textbf{Graphs $G$ with bounded arboricity}
In~\cite{ERS19-sidma, ERS20_soda}, Eden, Ron and Seshadhri first studied the problem of sublinear approximate counting in bounded arboricity graphs.
They presented improved algorithm for edges, star and $k$-clique counting in the general graph model, parameterized by the arboricity.
In~\cite{ERR19} presented an improved algorithm for almost uniform sampling of edges in bounded arboricity graphs, in the general graph query model.

\else

\bigskip
We review additional related work in Section~\ref{sec:related-work}, and next turn to discuss the ideas behind our upper and lower bounds.  
\fi

\subsection{The high level ideas behind the clique-sampling algorithm}\label{subsec:ub-overview}

We start by briefly describing the ideas behind the edge sampling algorithm of~\cite{ERR19} for sampling edges almost uniformly, which we employ both as a subroutine and starting point for our sampling scheme for $k$-cliques. Throughout, we assume that an upper bound $\alpha$ on the arboricity of the input graph is known.

\subsubsection{The edge sampling algorithm}\label{subsubsec:edg-samp-alg}

Let $L_0$ be the set of all vertices in the graph with degree at most (roughly) $\alpha$ (so that almost all the vertices in the graph belong to $L_0$).
The algorithm samples a vertex $v_0$ uniformly at random, and if $v_0$ is in $L_0$, it performs a short random walk
$v_0,v_1,\dots,v_j$ of length $j$ for $j$ chosen uniformly in $[\log n]$.
If at any point the walk returns to $L_0$ then the algorithm aborts, and otherwise, it returns the last edge traversed.

The analysis of the algorithm relies on a layered decomposition of the graph vertices. The vertices in $L_0$ comprise the first layer. Subsequent layers are defined inductively: a vertex $v$ is in $L_j$ if (1) it is not in any of the layers $L_i$ for $i<j$, and (2) most of its neighbors are in layers $L_0, L_2, \ldots, L_{j-1}$. While the algorithm is completely oblivious to the levels of the encountered vertices $v_i$ for $i>0$, using the aforementioned layering, it can be shown that each edge is sampled with almost equal probability $\approx \frac{1}{n\alpha}$.

\subsubsection{A random walk on $(k-1)$-cliques}

In order to sample $k$-cliques in $G$, we present a variant of the edge-sampling algorithm described in Section~\ref{subsubsec:edg-samp-alg},
and apply it to an auxiliary graph $H_k$, defined as follows. For each $(k-1)$-clique $Q$ in $G$, there is a node $v_{Q}$ in $H_k$, and for each $k$-clique $C$ in $G$, there is an edge in $H_k$ between a pair of nodes $v_{Q}, v_{Q'}$ corresponding to two of its $(k-1)$-cliques, $Q$ and $Q'$.
We say that $C$ is \emph{assigned} to $Q$ and $Q'$, where the assignment rule depends on the degrees (in $G$) of the vertices in $C$.
In particular, when $k=2$, this assignment is uniquely determined and we have $H_2=G$. In general, since there is a one-to-one correspondence between the edges in $H_k$ and the $k$-cliques of $G$, sampling an almost uniform edge in $H_k$ is equivalent to sampling an almost  uniform $k$-clique in $G$.
A simple but important observation is that if $G$ has arboricity at most $\alpha$, then so does $H_k$.

The challenge is to simulate the edge sampling procedure of~\cite{ERR19} on the graph $H_k$, while only having query access to the graph $G$. The simlation is not straightforward since:
\begin{compactenum}
\item we do not have query access to uniformly random nodes of $H_k$;
\item determining whether a node in $H_k$ is in layer $L_0$ can not be performed by a single degree query (as was the case in~\cite{ERR19}); and
\item in order to sample a random neighbor of a vertex $v_Q$ in $H_k$ we must sample a $k$-clique in $G$ that is assigned to $Q$. (In~\cite{ERR19} this could be acheived by a single neighbor query.)
\end{compactenum}
Below, we outline how we address 
these challenges.



\paragraph{Addressing the challenges.}

In order to sample nodes in $H_k$ we recursively invoke our procedure for sampling $(k-1)$-cliques in $G$ almost uniformly.
Given a sampled node $v_Q$ in $H_k$, we implement a procedure to check whether  $v_Q \in L_0$, by trying to approximate the number of $k$-cliques that are assigned to $Q$ in $G$. To do so efficiently, we replace the threshold $\alpha$ used to define $L_0$ in~\cite{ERR19}, by a value $\tau \geq \alpha$, where we will explain how $\tau$ is chosen later in the analysis.

It remains to explain how we simulate a random neighbor query for a vertex $v_Q$ in $H_k$ (so as to simulate a random walk on $H_k$).  Let $\mA(Q)$ denote the set of $k$-cliques assigned to $Q$. Recall that sampling an edge incident to $v_Q$ translates to sampling a $k$-clique $C$ in $\mA(Q)$. The rule for assigning each $k$-clique to a pair of $(k-1)$-(sub)cliques is defined in such a way that we need only consider $k$-cliques containing higher-degree neighbors of the minimum degree vertex $u$ in $V_Q$. Let $d(Q)=d(u)$, where $d(u)$ denotes the degree of $u$ in $G$. If $d(Q) \leq \sqrt{n\alpha}$, then by performing a uniformly random neighbor query from $u$ in $G$, each such neighbor is sampled with probability $1/d(Q)$. Otherwise if $d(Q)> \sqrt{n\alpha}$, then for every higher degree neighbor $w$ of $u$, $d(w)>\sqrt{n\alpha}$. By invoking the edge-sampling procedure of~\cite{ERR19} on $G$, we can sample every vertex $y$ in $G$ with probability $\frac{d(y)}{n\alpha}$. By performing rejection sampling, we can get each such higher degree vertex $w$ with probability $\frac{d(w)}{n\alpha}\cdot \frac{\sqrt{n\alpha}}{d(w)}=\frac{1}{\sqrt{n\alpha}}$.  Hence, for every $(k-1)$-clique $Q$, each $k$-clique in $\mA(Q)$ is sampled with probability $\max\left\{\frac{1}{d(Q)}, \frac{1}{\sqrt{n\alpha}}\right\} = \frac{1}{\min\{d(Q), \sqrt{n\alpha}\}}$.

To simulate a random neighbor query from a node $v_Q$ in $H_k$ such that $v_Q \notin L_0$ (so that the number of $k$-cliques in $\mA(Q)$ is at least $\tau$) we repeat the above sampling attempts $O^*\left( \left\lceil\frac{\min\{d(Q), \sqrt{n\alpha}\}}{\tau} \right\rceil\right)$ times. This process succeeds in obtaining a uniformly distributed $k$-clique in $\mA(Q)$ with high probability.
For a node $v_Q$ in $L_0$, performing $O^*\left( \left\lceil\frac{\min\{d(Q), \sqrt{n\alpha}\}}{\tau}\right\rceil\right)$ many attempts implies that each $k$-clique in $\mA(Q)$ is obtained with probability $1/\tau$.

An inductive analysis shows that a single invocation of the above simulation of the random walk on $H_k$ returns each $k$-clique in $G$ with probability roughly $\frac{1}{n\alpha \cdot \tau^{k-2}}$.
The $(n\alpha)$ term in the denominator comes from the base of the induction, i.e., sampling a uniform $2$-clique (edge) in $G$, and the term $\tau^{k-2}$ stems from the $k-2$ recursive calls, where in each level of recursion, we ``lose'' a factor of $1/\tau$.
Therefore, the overall success probability of a single attempt to sample an edge in $H_k$ is roughly $\frac{n_k}{n\alpha \cdot \tau^{k-2}}$. Hence, $O^*(\frac{n\alpha \cdot \tau^{k-2}}{n_k})$ repetitions are sufficient so that, with high probability, an almost uniformly distributed $k$-clique in $G$ is returned.
\tnote{change all refs to section to subsection}

\paragraph{Query complexity.}\label{sec:query_comp}
As discussed above, to sample a $k$-clique in $G$ with high probability, we perform $t=O^*\left(\frac{n\alpha \cdot \tau^{k-2}}{n_k}\right)$ repetitions of the random walk simulation on $H_k$. In each such simulation, there is a sequence of $k-1$ recursive calls to sample $i$-cliques for $i\in[2,\ldots,k]$ by performing a  random walk on the graph $H_i$.
Whenever a random neighbor query is simulated  on a node $v_T$ in $H_i$ for $i>2$,
$r=O^*\left(\frac{\min\{d(T), \sqrt{n\alpha}\}}{\tau}\right)$ queries are performed in $G$.
Conditioned on $\tau$ being sufficiently larger than $\alpha$, we get that the expected number of queries in each such simulation is just $O^*(1)$ (while the maximum is $O^*\left(\frac{\sqrt{n\alpha}}{\tau}\right)$). 
This implies that the  expected total query complexity is $O^*\left(\frac{n\alpha \cdot \tau^{k-2}}{n_k}\right)$.
As for the maximum running time, we can get an upper bound of $O^*\left(\frac{n\alpha \cdot \tau^{k-2}}{n_k}+ \frac{\sqrt{n\alpha}}{\tau}\right)$ by aborting the algorithm if it performs a larger number of queries, while still obtaining an output distribution as desired. Hence, we get a certain tradeoff between the expected query complexity and the maximum one (for ``hard to sample'' cliques).
In particular, if we set $\tau=\Theta^*(\alpha)$, we get that the expected query complexity is $O^*\left(
\frac{n\alpha^{k-1}}{n_k}\right)$, as in the case of counting, while the maximum query complexity is $O^*\left(
\frac{n\alpha^{k-1}}{n_k}+\sqrt{n/\alpha}\right).$
The upper bound in Theorem~\ref{thm:ub_informal} is derived by setting $\tau$ so that the two summands in the expression $O^*\left(\frac{n\alpha \cdot \tau^{k-2}}{n_k}+ \frac{\sqrt{n\alpha}}{\tau}\right)$ are equal.

\ifnum\fullversion=1
\paragraph{Where does the arboricity come into play.}\tnote{we can remove}\dnote{my tendency is not to include this here (maybe in a long version)}
We use the bound on the arboricity in two places. First, recall that the layer $L_0$ is defined according to the threshold $\tau$. Only for graphs with arboricity alpha and a setting of $\tau>\alpha/\eps$ it holds that graph decomposes into $\lceil \log n\rceil$ layers.
Second, recall that the expected cost of the simulations of uniform neighbor queries and verifying whether a node $v_T$ belongs to $L_0$ are bounded by $\left(\frac{d(T)}{\tau}\right)$. To bound the expected cost of this operation, we use a bound on the sum of degrees of all $k$-cliques in the graph, which is parameterized by the arboricity.
\fi

 \newcommand{\cG}{\mathcal{G}}
  \newcommand{\cP}{\mathcal{P}}
  \subsection{Overview of the lower bound}


\sloppy
The second term in the lower bound of Theorem~\ref{thm:lb} follows directly from a lower bound of
$\min\left\{n\alpha, \frac{n(\alpha/k)^{k-1}}{n_k}\right\}$
by~\cite{ERS20_soda} for the counting variant.
Hence, our main focus is on proving the first term (which, as noted previously, may be much larger than the second).

To obtain the first term in lower bound, we construct a family of graphs (with arboricity at most $\alpha$), such that in each graph, among the $n_k$ $k$-cliques that it contains, there is one ``hidden'' $k$-clique. This clique is hidden in the sense that 
any algorithm that (always) performs less than 
$\left(\frac{(n\alpha)^{k/2}}{k^k\cdot n_k}\right)^{\frac{1}{k-1}}/c$ (for a sufficiently large constant $c$) 
cannot  sample this clique with probability $\Omega(1/n_k)$.  This does not preclude the possibility that the expected complexity of the 
algorithm is smaller (as discusses in Subsection~\ref{sec:query_comp}). 

This is formalized by defining a process that answers the queries of a sampling algorithm ``on the fly'' while constructing a random graph in the family. All graphs in the family have the same underlying structure, and they differ in the choice of clique vertices and in the labeling of (part of) the edges. Here we give the high-level idea of the underlying structure, and the intuition for the lower-bound expression.

In each graph in the family, the hidden clique is over a subset $S$ of $k$ vertices that all have (high) degree $\Theta(\ell)$ where $\ell=\sqrt{n\alpha}$. The total number of high-degree vertices is $\Theta(\ell)$ as well. Other than the clique edges, there are no other edges between the high-degree vertices. Intuitively, in order to reveal the hidden clique, the algorithm must first reveal one edge $(u,u')$ in the clique and then reveal $k-2$ additional edges between $u$ and the other edges in the clique.\footnote{The algorithm may alternatively try to reveal $k/2$ edges in the clique that do not have common endpoints (or some other combination of edges that together are incident to all clique vertices), but this is not advantageous for the algorithm.} We prove that in each query, the probability of revealing the first edge of the clique is $O(k^2/\ell^2)$, and the probability of revealing any consecutive edge is $O(k/\ell)$.\footnote{We  note that whenever the term $\frac{(n\alpha)^{k/(2(k-1))}}{kn_k^{1/(k-1)}}$
 dominates the second term in the lower bound of Theorem~\ref{thm:lb},
it is smaller than $\sqrt{n/\alpha}$.}

The rough intuition for the upper bound $O(k^2/\ell^2)$ on revealing the first edge is that the number of clique edges is
$\binom{k}{2}$, while the total number of edges is $\Theta(\ell^2)$. Similarly, the rough intuition for the upper bound of $O(k/\ell)$ on revealing each additional edge in the clique is that each clique vertex has $k-1$ neighbors in the clique and a total of $\Theta(\ell)$ neighbors. In order to provide a formal argument, we
 define an auxiliary bipartite graph whose nodes correspond to graphs that are consistent with all previous queries (and answers)  and either contain a ``witness'' clique edge that corresponds to the query of the algorithm (one side of the graph), or do not (the other side). The edges of the bipartite graph are defined by certain transformations from witness graphs to non-witness graphs. By analyzing
the degrees of nodes on both sides of this auxiliary graph, we obtain the aforementioned bounds on the probability of revealing edges in the hidden clique.

Given these probability upper bounds, if an algorithm performs $T$ queries, then the probability that it reveals the hidden clique is upper bounded by  $T\cdot \frac{k^2}{\ell^2}\cdot \left(T\cdot \frac{k}{\ell}\right)^{k-2}$. If we want this expression to be $\Omega(1/n_k)$, the number of queries $T$ must be $\Omega\left(\left(\frac{(n\alpha)^{k/2}}{k^k\cdot  n_k}\right)^{1/(k-1)} \right)$.

\section{Preliminaries}\label{sec:prel}


\ifnum\fullversion=1
\begin{dfn}[Pointwise close to uniform.]
	Let $\mathcal{D}$ be a probability distribution on a set $\Omega$.
	We say that $\mathcal{D}$ is pointwise $\eps$-close to the  uniform distribution $\mathcal{U}$ on $\Omega$, if for every $x \in \Omega$,
	\[
	\Pr[x \sim \mathcal{D}]\in (1\pm\eps) \Pr[x \sim \mathcal{U}].
	\]
\end{dfn}

\begin{dfn}[Arboricity of a graph.]
	The arboricity of an undirected graph is the minimum number of forests into which its edges can be partitioned.
\end{dfn}
\fi

Let $G=(V, E)$ be a graph over $n$ vertices and arboricity at most $\alpha$.
Each vertex $v\in V$ has a unique id in $[n]$, denoted $id(v)$.
Let $\mC_k$ denote the set of $k$-cliques of $G$, and let $n_k=|\mC_k|.$ For a vertex $v$, let $\Gamma(v) = \Gamma_G(v)$ denote its set of neighbors and let $d(v)=d_G(v) = |\Gamma(v)|$.
We sometimes refer to edges as oriented, meaning that we consider each edge from both its endpoints.

Access to $G$ is given via the following types of queries:

\ifnum\fullversion=0
(1) A degree query, $\deg(v)$,  returns the degree $d(v)$ of the vertex $v$; (2) A neighbor queries, $\nbr(v,i)$ for $i\in[d(v)]$, returns the $i\th$ neighbor of $v$;
(3) A pair query, $\pair(v,v')$, returns whether $(v,v')\in E$.

\else
\begin{itemize}
	\item Degree queries, $\deg(v)$, that return the degree $d(v)$ of the vertex $v$.
	\item Neighbor queries $\nbr(v,i)$ that for $i\in[d(v)]$ return the $i\th$ neighbor of $v$.
	\item Pair queries $\pair(v,v')$ that return whether $(v,v')\in E$. \end{itemize}
\fi

\begin{dfn}[Ordering of the vertices.]
	We define an ordering on the graph's vertices, where $u\prec v$ if $d(u)<d(v)$ or if $d(u)=d(v)$	and $id(u)<id(v)$.
\end{dfn}

\begin{dfn}[Cliques' degree]
	For any  $k$-clique $C$, we let $d(C)$ denote the degree of the first vertex in $C$ according to the ordering $\prec$. We refer to it as the \emphdef{degree} of $C$. Letting $v$ denote the first vertex in $C$, we refer to $\Gamma(v)$ as the set of neighbors of $C$, and denote it by $\Gamma(C)$.
\end{dfn}

\begin{dfn}[Cliques id and an ordering of cliques]
	For a $t$ clique $C$,  let its id, $id(C)$ be a concatenation of its vertices ordered by $\prec$.
	We extend the order $\prec$ to cliques, so that for two $k$ cliques $C, C'$, $C\prec C'$ if $d(C)<d(C')$ or if
	$d(C)=d(C')$ and $id(C)<id(C')$.
\end{dfn}

\begin{dfn}[Assignment of $k$-cliques to $(k-1)$-cliques] \label{def:assign_clq}
	We \emph{assign} each $k$-clique $C$ its two first $(k-1)$-cliques according to $\prec$.
	For every $(k-1)$ clique $Q$, we denote its set of assigned $k$-cliques by $\mA(Q)$, and let $\dt(Q)=|\mA(Q)|$. We refer to $\dt(Q)$ as the \emph{assigned cliques degree} of $Q$.
\end{dfn}

\begin{obs}\label{clm:assign_clq_exceeds}\label{obs:dQ}
	Observe that by the above definition, if $Q$ and $Q'$ are assigned a $k$-clique $C$, then $d(Q)=d(Q')=d(C)$.
	Hence, if a $k$-clique $C$ is assigned to a  $(k-1)$-cliques $Q$ such that $C=Q\cup\{w\}$, then $d(Q)=d(C)\leq d(w)$.
\end{obs}
\noindent
We shall sometimes abuse notation and, let $\{Q,u\}$ denote $Q\cup \{w\}$.

\section{The algorithm for sampling $k$-cliques}\label{sec:ub}

As discuss in the introduction (Section~\ref{subsec:ub-overview}), our algorithm for sampling $k$-cliques almost uniformly in a graph $G$ works by simulating an edge-sampling procedure on an auxiliary graph, $H_k$, which is defined based on $G$. We begin  by precisely defining $H_k$, providing the clique-sampling algorithm and stating our main theorem.
In Section~\ref{subsec:sample-edge} we present  the edge-sampling procedure. This procedure is designed to run on a graph $F$, which is defined based on a graph $G$, where query access to $F$ is implemented by
subroutines that have query access to $G$. In our case the edge-sampling procedure is applied to $F=H_k$,
and as one of the subroutines requires access to almost uniformly distributed cliques of smaller size,
the procedure is also recursively applied to $F=H_i$, for $i =k-1, \ldots, 2$.
We give sufficient conditions on these subroutines based on which the correctness of the edge-sampling procedure can be established. We then turn (in Section~\ref{subsec:simulate-on-Hk}) to present and analyze these subroutines for
the $H_i$s.
We wrap things up with an inductive argument in Section~\ref{subsec:rec}.

\begin{dfn}[The graph $\Gt$]\label{def:Gt}\label{def:Hk}
	Given a graph $G$, we define the graph $\Gt(G)=\Gt=(\Vt, \Et)$ as follows. For every
	$(k-1)$-clique $Q$ in $G$ there is a node $v_{Q}$ in $\Vt$.
	For every $k$-clique $C$ in $ G$, there is an edge in $\Gt$ between the two $(k-1)$-cliques
	that $C$ is assigned to, as defined in Definition~\ref{def:assign_clq}.
\end{dfn}
For the sake of clarity, throughout the paper, we refer to the vertices
of $H_k$ as nodes.
Note that for the special case of $k=2$, we have that $\Gt=G$, and each edge ($2$-clique) in $G$, is assigned to both its endpoints.
More generally, the above definition implies a one-to-one correspondence between the set  edges incident to a node $v_{Q}$ in $\Gt$ and the set $\mA(Q)$ of $k$-cliques assigned to $Q$ in $G$.
Hence, sampling an edge $e\in \Gt$  is equivalent to sampling a $k$-clique in $G$. In the rest of the paper we go back and forth between sampling edges in $H_k$  and sampling $k$-cliques in $G$.

We claim that for every $k$, the graph $H_k(G)$ has arboricity at most $\alpha(G)$. The proof is deferred to Appendix~\ref{sec:appendix}.
\begin{clm}\label{clm:Hk_arb}
	Let $G$ be a graph of arboricity at most $\alpha$. Then $H_k(G)$
	has arboricity at most $\alpha$.
\end{clm}

In addition to receiving as input $n$, $\alpha$, $k$ and $\epsilon$ (as well as being given query access to $G$), our algorithm receives two additional parameters. The first, $\onk$, is assumed to be a constant factor estimate of $n_k$.
Such an estimate can be obtained by running the algorithm of~\cite{ERS20_soda} without asymptotically increasing the expected complexity of our sampling algorithm. The second, $\tau$, is a parameter that affects the complexity of our algorithm. In particular, as we show in Theorem~\ref{thm:ub-with-tau}, it introduces a certain tradeoff between the expected running time of the algorithm and the maximum running time.
We then show how for an appropriate setting of $\tau$ and by cutting-off the execution of the algorithm, we can obtain Theorem~\ref{thm:ub_informal}.

\newcommand{\setr}{\max \left\{ \frac{\sqrt{n\alpha}}{\tau},\min\left\{n\alpha, \frac{n\alpha\tau^{k-2}}{\onk} \right\}\right\} \cdot (k\log n/\eps)^{c k}}
\begin{figure}[htb!] \label{sclq}
	\fbox{
		\begin{minipage}{0.95\textwidth}
			\sclq$(G,n,\alpha,k,\eps,\onk,\tau)$
			\smallskip
			\begin{compactenum}
				\item Let  $\beta=\eps/10k$.  \label{step:set}
				\item Set $\overline{N}_{{k}}=n\alpha^{k-1}$.
				\item While the number of queries does not exceed $r=\setr$ for a sufficiently large constant $c$:\label{step:sclq_halt}
				\begin{compactenum}
					\item Invoke \se$(H_k,\overline{N}_{{k}},\beta ,\tau, \vec{p}=(G,n,k,\alpha))$, and if an edge in $H_k$ is returned, then \textbf{return} the corresponding $k$-clique in $G$.
				\end{compactenum}
			\end{compactenum}
		\end{minipage}
	}
\end{figure}

\begin{thm}\label{thm:ub-with-tau}
	Consider an invocation 	of the 
algorithm \sclq$(G,n,\alpha,k,\eps,\onk,\tau)$ with an estimate $\onk$ of $n_k$ such that $\onk \in [n_k, 2n_k]$, and  with a parameter $\tau$ such that $\frac{\alpha}{\eps}\cdot (4k\log n)^{k} \leq \tau\leq \sqrt{n\alpha}$.
	The
algorithm \sclq\ returns a $k$-clique $C$ in $\mC_k$, such that the resulting distribution on $k$-cliques is pointwise $\eps$-close to uniform on $\mC_k$.
	The expected query complexity of the algorithm is
$
O^*\left(  \min\left\{n\alpha,\frac{n\alpha\tau^{k-2}}{n_k} \right\}\right),$
	and the maximum query complexity of the
algorithm is
	$
	O^*\left(\max\left\{\frac{\sqrt{n\alpha}}{\tau}  ,\; \min\left\{n\alpha, \left(\frac{n\alpha\tau^{k-2}}{n_k}\right)
	\right\}\right\}\right).
	$
\end{thm}

Theorem~\ref{thm:ub_informal} is a corollary of the above theorem, when setting $\tau=\settau$, for a given good estimate $\onk$ of $n_k$.

\begin{cor}[Theorem~\ref{thm:ub_informal}, restated]\label{cor:ub}
	There exists a pointwise  $\eps$-close to uniform sampling algorithm for $k$-cliques in graphs with arboricity at most $\alpha$.
	If the algorithm is given an estimate $\onk$ of $n_k$ such that $\onk\in[n_k,2n_k]$, then the query complexity of the algorithm is
	\[
	O^*\left(\max\left\{ \left(\frac{(n\alpha)^{k/2}}{n_k}\right)^{\frac{1}{k-1}} ,\; \min\left\{n\alpha,\frac{n\alpha^{k-1}}{n_k} \right\}\right\}\right).
	\]
	If no such estimate is given, then the above only holds in expectation.
\end{cor}

\new{Note that for a given estimate $\onk$, we set 
$\tau=\settau$, while 
Theorem~\ref{thm:ub-with-tau} holds only for values $\tau$ such that 
$\frac{\alpha}{\eps}\cdot (4k\log n)^{k} \leq \tau\leq \sqrt{n\alpha}$. This setting implies that $\tau$ is always lower bounded by $\frac{\alpha}{\eps}\cdot (4k\log n)^{k}$, but is only upper bounded by $\sqrt{n\alpha}$ if $\alpha \leq \frac{\eps\sqrt{n\alpha}}{(4k\log n)^k}$. If this condition does not hold, then $\alpha=\tilde\Theta(n)$ (recall that always $\alpha\leq n$), so it is more beneficial to invoke the $m^{k/2}/n_k$ algorithm of~\cite{Peng}, replacing each edge query by an invocation of the edge sampling algorithm of~\cite{ERR19}.}

We defer the proof of the theorem and corollary to Subsection~\ref{subsec:rec}, and first present and analyze the procedure
\se\ and the subroutines it uses.

\subsection{The procedure \se}\label{subsec:sample-edge}

In this subsection we present the procedure \se\ for sampling edges almost uniformly in a graph $F$, given query access to a graph $G$ that defines $F$.
In the case that $F=G$, we simply invoke the procedure referred to in  the following theorem.
\begin{thm}[Corollary 2.8 in ~\cite{ERR19}.]\label{thm:ERR}\label{SBE}
	Let $G$ be a graph over $n$ vertices.
	There exists a procedure \SBE\ that, given query access to $G$, a bound $\alpha$ on the arboricity of $G$
	and a parameter $\beta \in (0,1)$, returns each oriented edge in $G$ with probability $\frac{1\pm \beta}{n\alpha \gamma}$ for $\gamma=4\log^2 n/\beta$ (and fails to return an edge with the remaining probability). The query complexity and running time of the algorithm are $O(\log n)$.
\end{thm}
For $F \neq G$, the procedure \se\ makes calls to three subroutines:  \ISLZ, \slz\  and \snbr.
These subroutines simulate queries to $F$ by performing queries to $G$. In particular,
 \ISLZ\ returns whether a given node in $F$ belongs to a certain set of nodes $L_0$ and effectively determines $L_0$ by its answers.
The subroutine \slz\ returns a random edge in $F$ that is incident to $L_0$
and, given a node $v$ in $F$ that does not belong to $L_0$, \snbr\ returns a random neighbor of $v$.
Using these subroutines, $\se$ performs a
random walk of length logarithmic in $|V_F|$ starting from $L_0$, and returns the last edge traversed in the walk.

\begin{figure}[htb!] \label{se}
	\fbox{
\begin{minipage}{0.95\textwidth}
	$\se(F, \overline{N}_F,\beta,\tau,\vec{p}=(G,n,\alpha,k))$
	\smallskip
	\begin{compactenum}
		\item If $F=G$, then invoke \SBE$(G,n,\beta,\vec{p})$ and if an edge $(u,v)$ is returned, then \textbf{return} it. Otherwise, \FAIL.
		\item Set $s=\lceil \log \oN_F\rceil$ and set $\beta'=\beta/(2s+2)$.		\label{step:se_setting}
		\item Choose $j \in [0,\ldots, s]$ uniformly at random. \label{step:se_j}
		\item Invoke \slz$( F, \oN_F,\beta',\tau,\vec{p})$,
		and let ${e}_0=(v_0,v_1)$ be the returned edge if one was returned. Otherwise, \textbf{return} \FAIL.   \label{step:sallow}
		\item For $i=1$ to $j$ do:  \label{step:se_loop}
		\begin{compactenum}
			\item \label{step:check_L0}			
			If \ISLZ$(F,\oN_F, v_i,\delta=\frac{\beta'}{\oN_F}\cdot (\beta/(k\log (n/\beta)))^{O(k)},\beta', \tau,\vec{p})$=YES then \textbf{return} FAIL. 	\label{step:se_ISLZ}
			\item Invoke \snbr$(F,v_i,\beta', \tau,\vec{p})$ to sample an edge $(v_i, v_{i+1})$ in $F$. \label{step:samp_nbr}
		\end{compactenum}	
		\item \textbf{Return} $(v_j,v_{j+1})$. 	
	\end{compactenum}
\end{minipage}
}
\end{figure}
\tnote{Get rid of $F$?}
\tnote{Change all $G$'s to $\mathcal{O}_G$ }

Before analyzing \se, we briefly discuss its application to sampling $k$-cliques.
As described in the algorithm \sclq,
in order to sample $k$-cliques in $G$, we 
invoke the procedure \se\ on the graph $H_k$ and implement the subroutines it calls (for $F=H_k$).
Our main lemma regarding the correctness and complexity of \se\ when invoked with $H_k$ 
(and for appropriate implementations of the subroutines) follows.
\begin{lem}\label{lem:main_lem}
	Consider an invocation of \se$(H_k, \overline{N}_{k},\beta ,\tau,\vec{p}=(G,n,k,\alpha))$ where $\overline{N}_{k}\geq|V_{H_k}|$,  and $(\alpha/\beta)\cdot O(4k\log n)^k \leq \tau\leq \sqrt{n\alpha}$.
	
The procedure returns an edge in $H_k$ with probability
\begin{equation*}
\Omega\left(
\frac{n_k}{n\alpha\gamma\cdot \tau^{k-2}}\cdot \left(\frac{\beta}{\log(n/\beta)}\right)^{O(k)}\;
\right),
\end{equation*}
and conditioned on an edge being returned, each edge is returned with probability $\frac{1\pm\beta}{n_k}$.

Furthermore, the expected running time of a single invocation of the procedure is $O^*(1)$, and the maximum running time is $O^*(\sqrt{n\alpha}/\tau)$.
\end{lem}

The proof of Lemma~\ref{lem:main_lem} appears in Section~\ref{subsec:rec}. There, we also prove Theorem~\ref{thm:ub-with-tau}, which follows almost directly from the lemma.

\medskip
\tnote{comment about goodness} 
\dnote{I would like to somehow say that they (in particular the edge sampling procedure) ``behave well'' (possibly under certain conditions) rather that ``they are good''. But I can't find a way to say it properly, so I think I'll live with the current phrasings of the definitions}
We next introduce several definitions that formalize what it means for the subroutines called by \se\ to be ``good simulators'' for queries to $F$.

\tnote{Change everything so that assume $L_0$ is good, and move failure probability analysis to sample-a-clq}
\begin{dfn}[A good $L_0$ and $L_0$ oracle]\label{def:good_L0}
	For a graph $F$ and a parameter $\tau$, we say that a subset of vertices in $F$ is a \emphdef{$\tau$-good $L_0$ with respect to $F$} if the following conditions hold: (1) for every vertex $v\in V_F$ such that $d_F(v)\leq \tau$ we have $v\in L_0$, and (2) for every $v\in V_F$ such that $d_F(v)>2\tau$ we have $v\notin L_0$.
	
	We say that a subroutine is a \emphdef{$(\delta,\tau)$-good $L_0$ oracle with respect to a graph $F$}, if with probability at least $1-\delta$ the subroutine defines a $\tau$-good $L_0$.
\end{dfn}

\begin{dfn}[A good $E(L_0)$-sampling subroutine.] \label{def:good_slz}
	We say that a subroutine is a \emphdef{$(\beta, X)$-good $E(L_0)$-sampling subroutine with respect to a graph $F$} it returns every oriented edge in $E_F(L_0)$ with probability in $\frac{1\pm\beta}{X}$.
\end{dfn}

\begin{dfn}[A good neighbor-sampling subroutine.] \label{def:good_snbr}
	We say that a subroutine is a \emphdef{$(\beta, \tau)$-good neighbor-sampling subroutine  with respect to a graph $F$}  if the following holds.
	Given a vertex $v$ such that $d_F(v)>\tau$,
		each oriented edge  incident to $v$ in $F$  is returned with probability  in $\frac{1\pm\beta}{d_F(v)}$.
\end{dfn}

In order to state and apply the next lemma, it will also be useful to introduce the following definition.
\begin{dfn}[A good edge-sampling procedure] \label{def:good_edge_sampling}
	We say that a procedure
is a \emphdef{$(\beta, X)$-good edge-sampling procedure
with respect to a graph $F$} if it returns every oriented edge in $F$ with probability in $\frac{1\pm\beta}{X}$.
\end{dfn}

Recall that $s=\lceil \log \oN_F\rceil$ and $\beta'=\beta/(2s+2)$.
\begin{lem}\label{lem:rw}
	Consider an invocation of the procedure \se\ with parameters $(F,\overline{N}_F, \beta, \tau,\vec{p})$
	Assume that the following conditions hold for some $X$.
	\begin{compactenum}
 		\item The subroutine \ISLZ$(F,\oN_F, v,\delta, \beta',\tau, \vec{p})$  is a $(\delta, \tau)$-good $L_0$ oracle with respect to $F$ for $\delta\leq \beta/X$.
		\item 	If $L_0$ determined by \ISLZ\ is $\tau$-good with respect to $F$, then
		\slz$(F,\oN_{F}, \beta', \tau,\vec{p})$ invoked in Step~\ref{step:sallow} is a $(\beta',X)$-good $E(L_0)$-sampling subroutine  for $L_0$.
		\item The subroutine \snbr$(F, v, \beta', \tau,\vec{p})$ invoked in Step~\ref{step:samp_nbr} is a $\beta'$-good neighbor-sampling subroutine for $F$.
	\end{compactenum}
	Then \se\ is a $(\beta,X')$-good edge-sampling procedure for $X' = X(s+1)$.
\end{lem}
The proof of Lemma~\ref{lem:rw} is similar to the proof of the correctness of the edge-sampling procedure of~\cite{ERR19}, with careful adaptations due to the simulated queries, and is deferred to Appendix~\ref{sec:proof_rw}.

We shall bound the complexity of \se\ in Section~\ref{subsec:rec}, after analyzing the subroutines.

\subsection{Simulating queries in $H_t$ for $t\leq k$: implementing the subroutines}\label{subsec:simulate-on-Hk}
In this section we present the
subroutines \ISLZ, \slz\ and \snbr.
The subroutines are stated for $H_k$, but can be applied to any $H_t$ for $t\in[3,k]$.
\tnote{change everything to $H_t$}

We will rely on the following simple claims from~\cite{ERS20_soda}.

\begin{clm}[Claim 3.1 in~\cite{ERS20_soda}]\label{clm:bound-clq-deg}
	For every $k\geq 2$,
	$ \sum_{C \in \mC_k} d(C)\leq n\cdot \alpha(G)^{k}.$
\end{clm}

\begin{clm}[Claim 3.2 in~\cite{ERS20_soda}]\label{clm:bound-clqs}
	For  every $k\geq 2$,
	$n_k(G) \leq \frac{2\alpha(G)}{k}\cdot n_{k-1(G)}$.
\end{clm}

The next simple claim is useful for estimating the probability of the recurring sampling attempts in the different
subroutines.
The proof is deferred to Appendix~\ref{proof:taylor}.
\begin{clm}\label{clm:taylor}
	For $x>0$, $y\geq 1$ and $x\cdot \lceil y\rceil <1$, it holds that
	\[
	xy(1-2xy)\leq 1-(1-x)^{\lceil y \rceil} \leq xy(1+1/y).
	\]
\end{clm}

We start with presenting and analyzing the procedure for sampling
oriented edges incident to nodes in $L_0(\Gt)$ (where $L_0$ will be determined by the subroutine \ISLZ\ as discussed subsequently).
Observe that this is the subroutine in which the recursive calls are invoked. Indeed, to sample edges in $E(L_0(\Gt))$, it first samples (almost uniformly distributed) nodes in $\Gt$, which is equivalent to sampling  (almost uniformly distributed) $(k-1)$-clique in $G$. Therefore, in this subsection we analyze the correctness conditioned on the correctness of the recursive call to \se, and in the following section we shall prove the complete inductive argument.

\begin{figure}[htb!] \label{slz}
\fbox{
\begin{minipage}{0.95\textwidth}
	$\slz(\Gt,\beta', \tau,\vec{p}=(G,n,k,\alpha))$
	\smallskip
	\begin{compactenum}
		\item Let $\overline{N}_{k-1}=n\alpha^{k-3}$.
		\item Let $\beta=\beta'/4$.
		\item Invoke \se$(H_{k-1},\overline{N}_{k-1},\beta,\tau,\vec{p})$. Let $Q$ be the returned $(k-1)$-clique if one was returned. Otherwise \FAIL. \label{step:slz_rec}
		\item If \ISLZ$(\Gt, v_Q,\delta=\frac{\beta'}{\oN_F}\cdot (\beta/(k\log (n/\beta)))^{O(k)},\beta',\tau,\vec{p})$=NO  then \FAIL.\label{step:slz_check_L0}
		\item If $d(Q) \leq 40\tau/\beta^2$, then flip a coin with bias $d(Q)/(40\tau/\beta^2)$, and if it comes out \textsf{Heads}	:
		\begin{compactenum}
			\item Select a neighbor $w$ of $Q$ uniformly at random.
		\end{compactenum}
		\item Else, flip a coin with bias $\beta^2/40$ and if the coin turns out \textsf{Heads} then
		repeat the following at most $\left\lceil \frac{3\min\{d(Q),\sqrt{n\alpha\gamma}\}}{\beta\tau} \right\rceil$ times: \label{step:slz_repeat_inner}
		\begin{compactenum}
			\item If  $d(Q) \leq \sqrt{n\alpha\gamma}$, then:
			\begin{compactenum}               	
				\item Select a neighbor $w$ of $Q$ uniformly at random.
			\end{compactenum}
			\item Otherwise ($d(Q) > \sqrt{n\alpha\gamma}$):
			\begin{compactenum}
				\item Invoke the procedure \SBE$(G,n,\beta/3,\vec{p})$. Let $(w,z)$ denote the returned edge if one was returned.
				\item If $d(w)\geq d(Q)$, then keep $w$ with probability $\sqrt{n\alpha\gamma}/d(w)$.
			\end{compactenum}
			\item Check if $\{Q,w\}$ is a $k$-clique assigned to $Q$. If so, \textbf{return} $C=\{Q,w\}$.
		\end{compactenum}
	\end{compactenum}
	
\end{minipage}
}
\end{figure}

\begin{clm}\label{clm:slz}
	Consider an invocation of \slz\ with  $\Gt, \beta,\tau$  and $\vec{p}=(G,n,k,\alpha)$ such that  $\frac{4\alpha}{\beta} \leq \tau\leq \sqrt{n\alpha}$.
	Assume that
	$L_0$ is $\tau$-good, and that 
	\se\ invoked in Step~\ref{step:slz_rec} is a $(\beta,X)$-good edge-sampling procedure with respect to $H_{k-1}$ for some $X$ such that $X\tau^2\geq  n\alpha^{k}$. 	
	
	Then the subroutine \slz is a $(\beta',X')$-good $E(L_0)$-sampling subroutine with respect to $\Gt$  for $X'=40X\tau/\beta^2$ and $\beta'=4\beta$ (recall Definition~\ref{def:good_slz}).
	
	Furthermore, the expected query and time complexity of the subroutine are $O(q_1+\log n)$ and the maximum query and time complexity are $O(q_2 + \frac{\log n \sqrt{n\alpha\gamma}}{\tau})$, where $q_1$ and $q_2$ are the expected and maximum, respectively, query and time complexity of \se$(H_{k-1}, \overline{N}_{{k-1}}, \beta,\tau,\vec{p})$.
\end{clm}
\begin{proof}
	Due to Step~\ref{step:slz_check_L0}, if in Step~\ref{step:slz_rec} a $(k-1)$-clique $Q$ is sampled  such that  $v_{Q}\notin L_0(\Gt)$, then the subroutine fails. Hence, only edges in $E(L_0(F))$ have non-zero probability of being returned.
	Consider an oriented edge $(v_{Q},v_{Q'})$ for some $v_Q \in L_0(\Gt)$.
	By the definition of $\Gt$, the edge $(v_Q,v_{Q'})$ in $\Gt$ corresponds to a $k$-clique $C$ in $G$ that is assigned to the $(k-1)$-clique $Q$.
	By the assumption that \sclq\ is a $(\beta,X)$-good $(k-1)$-clique sampling subroutine, $Q$ is sampled in Step~\ref{step:slz_rec} with probability $\frac{1\pm\beta}{X}.$
	Condition on this event happening.
	
	First consider the case that  $d(Q)\leq \frac{40\tau}{\beta^2}$. Then
	$w$ is sampled with probability $\frac{d(Q)}{40\tau/\beta^2}\cdot \frac{1}{d(Q)}=\frac{\beta^2}{40\tau}$.
	
	Next consider the case that  $d(Q)>40\tau/\beta^2$. Note that in this case an edge is sampled only if the
	coin toss at Step~\ref{step:slz_repeat_inner} \textsf{Heads}.
	We again consider two separate cases.
	If $Q$ is such that $d(Q)\leq \sqrt{n\alpha\gamma}$, then every  clique $\{Q,w\} \in \mA(Q)$ is sampled with probability $\frac{\beta^2}{40d(Q)}$, where the $\beta^2/40$ factor is due to the tossing of the coin.
	Otherwise  $d(Q)>\sqrt{n\alpha\gamma}$, and by Claim~\ref{clm:assign_clq_exceeds}, $d(w)\geq d(C')>\sqrt{n\alpha\gamma}$.
	Furthermore, by Theorem~\ref{thm:ERR}, an invocation of Sample-an-Edge$(G,\alpha,\beta/3)$ returns each oriented edge in the graph with probability $\frac{1\pm\beta/3}{n\alpha\gamma}$ for $\gamma=\setGamma$.
	Hence, each assigned clique $\{Q,w\}$ in $\mA(Q)$ is sampled and kept with probability $\frac{(1\pm\beta/3)\beta d(w)}{n\alpha\gamma}\cdot \frac{\sqrt{n\alpha\gamma}}{d(w)}=\frac{(1\pm\beta/3)\cdot \beta^2}{40\sqrt{n\alpha\gamma}}$.
	It follows that for every $Q$ such that $d(Q)>40\tau/\beta^2$, in each invocation of the  loop in Step~\ref{step:slz_repeat_inner}, every $k$-clique in $\mA(Q)$ is sampled with almost equal probability $\frac{(1\pm\beta/3)\cdot\beta^2}{40\min\{ d(Q), \sqrt{n\alpha\gamma}\}}$.
	Hence,the success probability of a single invocation is $\frac{(1\pm\beta/3)\cdot\beta^2\dt(Q)}{40\min\{ d(Q), \sqrt{n\alpha\gamma}\}}$. Let $x$ denote this probability.

	Let   $y\eqdef \frac{3\min\{d(Q), \sqrt{n\alpha\gamma}\}}{\beta \tau} $, and $\mathcal{E}$ denote the event that a $k$-clique is returned in one of the invocations. Then
	\[\Pr[\mathcal{E}] = 1-(1-x)^{\lceil y \rceil} \approx 1-\left(1-\frac{(1\pm\beta/3)\cdot\beta^2\dt(Q)}{40\min\{ d(Q), \sqrt{n\alpha\gamma}\}}\right)^{\left\lceil \frac{3\min\{d(Q), \sqrt{n\alpha\gamma}\}}{\beta \tau} \right \rceil}.\]
	By the assumption that $L_0$ is $\tau$-good, and since $v_Q \in L_0(H_k)$, it holds that $\dt(Q)\leq 2\tau$, and thus  $x\lceil y \rceil <\frac{6(1+\beta/3)\cdot\beta\dt(Q)}{40\tau}< 1$. Therefore, 	by Claim~\ref{clm:taylor},   it holds that \tnote{verify}
	\[
	xy(1-2xy)\leq \Pr[\mathcal{E}]\leq xy(1+1/y).
	\]
	Since $\dt(Q)<2\tau$ and $\beta<1/2$,
	$2xy \leq\frac{6(1+\beta/3)\beta \dt(Q)}{40\tau} <\beta/3$.
	Also, since $d(Q)\geq 40\tau/\beta^2$, and by the assumption $\tau<\sqrt{n\alpha\gamma}$, it holds that
	$1/y=\frac{\beta \tau}{3\min\{d(Q), \sqrt{n\alpha\gamma}\}} \leq \beta/3$.
	Hence, $\Pr[\mathcal{E}]\in xy\cdot (1\pm\beta/3),$ implying
	\[\Pr[\mathcal{E}] \in \frac{(1\pm \beta/3) \beta\cdot \dt(Q)}{40\tau} \in \frac{(1\pm \beta) \cdot \beta \dt(Q)}{40\tau}\;.
	\]
	Also, observe that conditioned on  a $k$-clique being  returned, each $k$-clique is returned with probability $\frac{1}{(1\pm\beta/3)\dt(Q)}$. 
	Therefore, for the case  $d(Q)>40\tau/\beta^2$, it holds that for every specific assigned $k$-clique $C$ of $Q$,
	\[\Pr[C \text{ is returned}]=\frac{(1\pm \beta/3)^2 \cdot \beta^2 \dt(Q)}{40\tau} \cdot \frac{1}{(1\pm\beta/3)\dt(Q)}\in(1\pm \beta) \cdot \frac{\beta^2}{40\tau}.
	\]	
	
	It follows that  for every $k$-clique $C$ corresponding to an oriented edge in $E(L_0(H_k))$,
	\[\Pr[C \text{ is returned}] \in \frac{1}{X}\cdot \frac{(1\pm\beta)\beta^2}{40\tau}=\frac{(1\pm\beta)\beta^2}{40X\tau} \]
	(independently of the degree of the clique to which it is assigned).
	
	We turn to analyze the query and time complexity of the procedure, which we shall denote by cost.
	Recall $q_1$ and $q_2$ denote the expected and maximum cost of a single invocation of the subroutine  \slz, respectively.
	Further note that by Theorem~\ref{thm:ERR}, each invocation of the procedure \SBE, takes $O(\log n)$ queries. Hence,
	\begin{align*}
	\EX[cost]&=O\left(q_1+\frac{1}{X\tau}\left(\sum_{\substack{Q\in \mC_{k-1} \text{ s.t.} \\ d(Q)\leq 40\tau/\beta^2}}\frac{\beta^2 \cdot d(Q)}{40\tau}\cdot q_1 + \sum_{\substack{Q\in \mC_{k-1} \text{ s.t.} \\ d(Q)>40\tau/\beta^2}} \beta^2\log n \cdot \left\lceil\frac{\min\{d(Q), \sqrt{n\alpha\gamma}\}}{40\tau\beta}\right\rceil \right)\right)
	\\& =O\left(q_1+ \frac{\beta\log n }{X\tau^2}  \sum_{Q'\in \mC_{k-1}} d(Q') \right)=O\left( q_1+\frac{\beta\log n \cdot n\alpha^{k-1}}{X\tau^2}\right)=O( q_1 \log n).
	\end{align*}
	where the second to last equality is due to Claim~\ref{clm:bound-clq-deg},
	and the last equality is by the assumption that $X\tau^2\geq n\alpha^{k-1}$.
	Also,  the maximum query complexity of the procedure is $O\left(q_2+\frac{\log n\sqrt{n\alpha\gamma}}{\tau} \right).$

\end{proof}

\renewcommand{\setr}{\frac{\min\{d(Q), \sqrt{n\alpha\gamma}\}}{\tau}\cdot \log(1/\beta')}
\begin{figure}[htb!] \label{snbr}
	\fbox{
		\begin{minipage}{0.95\textwidth}
			$\snbr(H_k, v_Q,\beta,\tau,\vec{p}=(G,n,k,\alpha))$
			\smallskip
			\begin{compactenum}				
				\item Let $Q$ be the $(k-1)$-clique in $G$ corresponding to $v_Q$.
				\item Repeat at most $r=\frac{\min\{d(Q), \sqrt{n\alpha\gamma}\}}{\tau}\cdot 2\ln(3/\beta)$ times:\label{step:snbr_loop}
				\begin{compactenum}
					\item If $d(Q)\leq \sqrt{n\alpha\gamma}$:
					\begin{compactenum}
						\item Sample a neighbor $w$ of $Q$ uniformly at random.
					\end{compactenum}
					\item Else, if $d(Q)>\sqrt{n\alpha\gamma}$:
					\begin{compactenum}
						\item Invoke the procedure \SBE$(G, n,\beta/3,\vec{p})$. Let $(w,z)$ denote the returned edge, if one was returned. Otherwise, return to Step~\ref{step:snbr_loop}. \label{step:snbr_be}
						\item If $d(w)\geq d(Q)$, then keep $w$ with probability $\sqrt{n\alpha\gamma}/d(w)$.
						Otherwise, return to Step~\ref{step:snbr_loop}.
					\end{compactenum}
					\item If $Q$ and $w$ form a $k$-clique $C$, and $C$ is assigned to $Q$,  then \textbf{return} the edge  in $F$ that corresponds to $C$.
				\end{compactenum}
			\end{compactenum}
		\end{minipage}
	}
\end{figure}

\begin{clm}\label{clm:snbr}
	The subroutine \snbr\ is a $(\beta,\tau)$-good neighbor-sampling subroutine with respect to $H_k$, as defined in Definition~\ref{def:good_snbr}.
	The query and time complexity of the subroutine are $O\left(\frac{\min\{d(Q), \sqrt{n\alpha\gamma}\}}{\tau}\cdot  \log n\log(1/\beta)\right)$.
\end{clm}
\begin{proof}
	Let $v_Q$ be a node in $\Gt$ such that $v_Q\notin L_0(\Gt)$.
	We shall prove that every $k$-clique in $\mA(Q)$ is returned with probability in $\frac{1\pm\beta}{\dt(Q)}$, implying that
	every 	incident edge of $v_Q$ in $\Gt$ is returned with probability $\frac{1\pm \beta}{d_{\Gt}(v)}$.
	
	If $d(Q)\leq \sqrt{n\alpha\gamma}$, then for every assigned clique  $C=\{Q,w\}$ of $Q$, $w$ is sampled with probability $1/d(Q)$.
	Otherwise $d(Q)>\sqrt{n\alpha\gamma}$, and  by Claim~\ref{clm:assign_clq_exceeds}, $d(w)\geq d(C)>\sqrt{n\alpha\gamma}$.
	Also, by Theorem~\ref{thm:ERR}, the invocation of \SBE$(G,n,\alpha,\beta/10, \vec{p})$ in Step~\ref{step:snbr_be} returns each oriented edge in the graph $G$ with probability $\frac{1\pm\beta/10}{n\alpha\gamma}$.
	Hence, each assigned clique $C=\{Q,w\}$ in $\mA(Q)$ is sampled and kept with probability $\frac{(1\pm\beta/10)d(w)}{n\alpha\gamma}\cdot \frac{\sqrt{n\alpha\gamma}}{d(w)}=\frac{1\pm\beta/10}{\sqrt{n\alpha\gamma}}$.
	Hence, for every $Q$ (independently of its degree), each $k$-clique assigned to $Q$ is sampled with probability
	$ 	\frac{1\pm \beta/10}{\min\{d(Q), \sqrt{n\alpha\gamma}\}}$.

	Let $\mathcal{E}$ denote the event that an assigned $k$-clique of $Q$ is sampled in one of the $r$ invocations of the loop in Step~\ref{step:snbr_loop}.
	Recall that by the assumptions that $v\notin L_0$ and that $L_0$ is $\tau$-good, it holds that $\dt(Q)>\tau$.
	Hence, for $r=\frac{\min\{d(Q), \sqrt{n\alpha\gamma}\}}{\tau}\cdot \ln(10/\beta)$,
	\begin{align*}
	\Pr[\neg \mathcal{E}]<  \left(1-\frac{(1-\beta/10)\dt(Q)}{\min\{d(Q), \sqrt{n\alpha\gamma}\}}\right)^r< \left(1-\frac{(1-\beta/10)\tau}{\min\{d(Q), \sqrt{n\alpha\gamma}\}}\right)^r <\beta/10 \;. \label{eqn:snbr_2}
	\end{align*}
	Also note that in each sampling attempt, every assigned $k$-clique has equal probability (up to a $(1\pm\beta/10)$ multiplicative factor) to be the returned one. Hence, conditioned on a $k$-clique being  returned, each is returned with probability $\frac{1\pm\beta/10}{(1\pm\beta/10)\dt(Q)}\in \frac{1\pm\beta/3}{\dt(Q)}$.
	Therefore, each $k$-clique in $\mA(Q)$ is returned with probability
	\[ \Pr[\mathcal{E}]\cdot \frac{1\pm\beta/3}{\dt(Q)}\in \frac{1\pm\beta}{\dt(Q)}.
	\]

	By Theorem~\ref{thm:ERR}, each invocation of \SBE\ has query and time complexity $O(\log n)$.
	There are at most $O(r)$ invocations of the procedure, and all other steps have query and time complexity $O(1)$. Hence,
	the query complexity and running time of the procedure are $O(r\cdot\log n)=O\left(\frac{\min\{d(Q), \sqrt{n\alpha\gamma}\}}{\tau}\cdot  \log n\log(1/\beta)\right)$.
	
\end{proof}

We now present our subroutine for determining  $L_0$.
Namely, $L_0$ is determined  according to the output of the subroutine, so that $L_0=\{v_Q\in V_{H_k} \mid \ISLZ(H_k,\oN_{k}, v_Q,\delta, \beta,\tau,\vec{p})=1 \}$ (where we assume that the randomness of the subroutine is uniquely determined for each $v_Q$).
Hence, \ISLZ\ determines $L_0$, and we would like to prove that the $L_0$ it determines is a $(\delta,\tau)$-good $L_0$ with respect to $H_{k}$ (recall Definition~\ref{def:good_L0}).

\renewcommand{\setr}{\frac{\min\{d(Q), \sqrt{n\alpha\gamma}\}}{\tau}\cdot 50\log(1/\delta')}
\begin{figure}[htb!] \label{ISLZ}
	\fbox{
		\begin{minipage}{0.95\textwidth}
			$\ISLZ(H_k,\oN_{k}, v_Q,\delta, \beta,\tau,\vec{p}=(G,n,k,\alpha))$
			\smallskip
			\begin{compactenum}				
				\item Let $Q$ be the $(k-1)$-clique in $G$ corresponding to $v_Q$.
				\item Let $\delta'=\delta/\oN_{k}$.
				\item For $i=1$ to $r=\setr$ times:\label{step:islz_loop}
				\begin{compactenum}
					\item If $d(Q)\leq \sqrt{n\alpha\gamma}$:
					\begin{compactenum}
						\item Sample a neighbor $w_i$ of $Q$ uniformly at random.
					\end{compactenum}
					\item Else, if $d(Q)>\sqrt{n\alpha\gamma}$:
					\begin{compactenum}
						\item Invoke the procedure \SBE$(G, n,\beta,\vec{p})$, and let $(w,z)$ denote the returned edge, if one was returned. 						Otherwise, return to Step~\ref{step:snbr_loop}.
						\item If $d(w_i)>d(Q)$, then keep $w_i$ with probability $\sqrt{n\alpha\gamma}/d(w_i)$.
						Otherwise, return to Step~\ref{step:snbr_loop}.
					\end{compactenum}
					\item If $Q$ and $w_i$ form a $k$-clique $C$, and $C$ is assigned to $Q$,  then let $\chi_{w_i}=1.$
				\end{compactenum}
				\item Let $\tilde{a}=\frac{1}{r}\sum_{i=1}^r\chi_{w_i}$.
				\item If $\tilde{a}<1.5 \tau/d(Q)$ then return YES. Otherwise, return NO.
			\end{compactenum}
		\end{minipage}
	}
\end{figure}
We note that if \ISLZ\ is invoked more than once with the same node $v_Q$, then it returns the same answer as in the first invocation with $v_Q$.

\begin{clm}\label{clm:ISLZ}
	Consider an invocation of the subroutine \ISLZ\ with parameters $H_{k}, \oN_k, v_Q,\delta,\beta,\tau$ and $\vec{p}=(G,n,k,\alpha)$.
	Let $Q$ denote the $(k-1)$-clique corresponding to the node $v_Q$.
	With probability at least $1-\delta/\oN_k$, the followings hold.
	\begin{itemize}
		\item If $\dt(Q)\leq \tau$, then the subroutine returns YES.
		\item  If $\dt(Q)>2\tau$, then the subroutine returns NO.
	\end{itemize}
	Furthermore, the query and time complexity of the subroutine are $O\left(\frac{\min\{d(Q), \sqrt{n\alpha\gamma}\}\cdot k\log^2(n/\delta)}{\tau}\right)$.
\end{clm}
\begin{proof}
	If  $d(Q) \leq \sqrt{n\alpha\gamma}$, then
	\[
	\EX_{w \in \Gamma(Q)}[\chi_{w}]=\frac{1}{d(Q)}\sum_{w\in \Gamma(w)} \mathbbm{1}_{\{Q,w\}\in \mA(Q)} =\frac{\dt(Q)}{d(Q)}.
	\]
	Otherwise, if $d(Q) > \sqrt{n\alpha\gamma}$, then by Claim~\ref{clm:assign_clq_exceeds}, for every $w$ such that $\{Q,w\}\in \mA(Q)$, $d(w)\geq d(Q)>\sqrt{n\alpha\gamma}$. 	By  Theorem~\ref{thm:ERR}, an invocation of \SBE$(G,n,\beta,\vec{p})$ returns each oriented edge in $G$ with probability $\frac{1\pm\beta}{n\alpha\gamma}$.
	Hence, each assigned clique $\{Q,w\}$ of $Q$ is sampled and kept with probability $\frac{(1\pm\beta)d(w)}{n\alpha\gamma}\cdot \frac{\sqrt{n\alpha\gamma}}{d(w)}=\frac{1\pm\beta}{\sqrt{n\alpha\gamma}}$.
	It follows that in the case that $d(Q) > \sqrt{n\alpha\gamma}$,
	\[
	\EX_{w \in \Gamma(Q)}[\chi_{w}]=\frac{1}{d(Q)}\sum_{w\in \Gamma(w)} \mathbbm{1}_{\{Q,w\}\in \mA(Q)} =\frac{(1\pm\beta)\dt(Q)}{\sqrt{n\alpha\gamma}}.
	\]
	Hence, in general, $\EX[\chi_w]\in\frac{(1\pm\beta)\dt(Q)}{\min\{d(Q),\sqrt{n\alpha\gamma}\}}$.
	Therefore, by the multiplicative Chernoff bound, as the $\chi_w$ variables are $\{0,1\}$ independent variables, it holds that
	\[
	\Pr\left[\left|\frac{1}{r}\sum_{i=1}^r \chi_{w_i}-\EX[\chi_w]\right|>\frac{1}{4}\EX[\chi_w]| \right]<\exp\left(-\frac{r\cdot \EX[\chi_w]}{48}\right)\;. \label{eq:chernoff}
	\]
	
	For $Q$ such that $\dt(Q)>2\tau$, $\EX[\chi_w]>\frac{(1-\beta)\cdot 2\tau}{d(Q)}$, so that
	by Equation~\eqref{eq:chernoff} and the setting of $r=\setr$,
	$\Pr[\tilde{a}<\frac{3}{4}\cdot \frac{(1-\beta)2\tau}{d(Q)}]<\delta'$.
	Hence, with probability at least $1-\delta'$, $\tilde{a}>1.5\frac{\tau}{d(Q)}$ and the subroutine will return NO (where we used $\beta<1/4$).
	
	If $d(Q)<\tau$, then consider a random variable $\chi'$ for which $\EX[\chi']=\frac{(1+\beta)\tau}{d(Q)}$.
	Then, by Equation~\eqref{eq:chernoff} and the setting of $r$ and $\tilde{a}$,
	\[\Pr\left[ \frac{1}{r}\sum_{i=1}^r\chi_{w_i} >\frac{5}{4}\cdot \frac{(1+\beta)\tau}{d(Q)}\right]\leq \Pr\left[ \frac{1}{r}\sum_{i=1}^r\chi'_i >\frac{5}{4}\cdot \frac{(1+\beta)\tau}{d(Q)}\right]<\delta'.
	\]
	Hence, with probability at least $1-\delta'$, $\tilde{a}< 1.5\tau/d(Q)$ the subroutine will return YES (where here too we used $\beta<1/4$).
	
	Finally, by Theorem~\ref{thm:ERR}, the query complexity of each invocation of the loop is $O(\log n)$ in case \SBE\ is invoked, and  $O(1)$ otherwise.
	By the setting of $\delta'=\delta/\oN_F$, it follows that  the query and time complexity of the subroutine are $O(r)=O\left(\frac{\min\{d(Q), \sqrt{n\alpha\gamma}\}\cdot \log ^2(n/\delta)}{\tau}\right)$ as claimed.	
\end{proof}

\begin{cor}\label{cor:ISLZ_oracle}
	If the  subroutine \ISLZ\ is invoked with $H_k,\oN_k,\delta,\beta,\tau$ such that $|H_k|\leq \oN_k$, then
	it is a $(\delta, \tau)$-good $L_0$ oracle with respect to $H_k$, as defined in Definition~\ref{def:good_L0}.
	That is, with probability at least $1-\delta$, the subroutine \ISLZ\ defines a $\tau$-good set $L_0$ for the graph $H_k$: for every $v_Q$ such that $\dt(Q)\leq \tau$, it returns YES, and for every $(k-1)$-clique $Q$ such that $\dt(Q)>2\tau$, it returns NO.
\end{cor}
\begin{proof}
	Assume that each $(k-1)$-clique $Q$ is assigned a unique random string $R_Q$ so that whenever \ISLZ\ is invoked with $v_Q$ the subroutine uses the same randomness $R_Q$. Then we get that the with probability at least $1-\delta$,  over the choice of these random strings, all $Q$ cliques are correctly classified with probability at least $1-\oN_k\cdot \frac{\delta}{\oN_k}>1-\delta$. Hence, with probability at least $1-\delta$, \ISLZ\ is an oracle to a $\tau$-good set $L_0$.
\end{proof}

\subsection{Proving the recursion}\label{subsec:rec}

Given the statements regarding the query complexity of the 
subroutines
\slz\, \snbr\ and \ISLZ, we are now ready to prove the statement regarding the query complexity of the procedure \se.

\begin{lem}[\se\ complexity]\label{clm:se_comp}
	Consider an invocation of the procedure \se\ with parameters $(F,\overline{N}_F, \beta, \vec{p}=(G,n, k, \tau,\alpha))$.
	Let $q^{exp}_{n,\tau,\gamma}(k-1,\beta')$ and 	$q^{max}_{n,\tau,\gamma}(k-1,\beta')$
denotes the expected and maximum query complexities of the recursive invocation of the procedure \se\ in Step~\ref{step:slz_rec} of \slz.
Also, 	assume that the followings hold.
	\begin{compactenum}
		\item 	if \ISLZ\ is $\tau$-good with respect to $H_{k}$ then 
		\slz$(H_k,\oN_{k}, \beta', \vec{p})$ invoked in Step~\ref{step:sallow} is a $(\beta',X)$-good $E(L_0)$-sampling subroutine with respect to $H_k$ for $X$ such that $X\cdot \tau=\Omega^*(n\alpha^{k})$.
		\item 	\ISLZ\ invoked in Step~\ref{step:check_L0} is a $(\delta, \tau)$-good $L_0$-oracle with respect to $H_k$ for $\delta\leq \beta'/X$.
		\item \snbr$(H_k, v_Q, \beta',\vec{p})$ invoked in Step~\ref{step:samp_nbr} is a $\beta'$-good neighbor-sampling subroutine with respect to $H_k$.
	\end{compactenum}		
	Then for $k=2$ 	the expected and maximum query and time complexities of the procedure are $O(\log n)$.
	For $k\geq 3$, 	
	the expected query and time complexity of the procedure is
	 $O^*\left(q^{exp}_{n,\tau,\gamma}(k-1,\beta/(2s+2))	+1 \right)$,
	and the maximum query complexity is
	$O^*\left(q^{max}_{n,\tau,\gamma}(k-1,\beta/(2s+2)) + \frac{\sqrt{n\alpha}}{\tau}\right)$.
\end{lem}
\begin{proof}
	First, if $k=2$, then observe that the procedure simply invokes \SBE, and returns an edge if one was returned. 
	By Theorem~\ref{thm:ERR}, the invocation of \SBE\ takes $O(\log n)$ queries, and the claim holds.
	Hence, assume $k\geq 3$.
	
	We start by analyzing  the expected cost of each invocation of the loop in Step~\ref{step:se_loop}.
	Let $e=(v_{Q_i}, v_{Q_{i+1}})$ denote the $i\th$ traversed edge in the random walk, and let $C$ denote the corresponding $k$-clique to $e$.
	
	By Claim~\ref{clm:ISLZ}, since \ISLZ\ is invoked with $\delta=(\beta'/\oN_{k-1})\cdot (\beta/(k\log (n/\beta)))^{O(k)}$ and $\oN_{k-1}=n\alpha^{k-2}$, the complexity of Step~\ref{step:samp_nbr} is 
	$O^*\left(\frac{\min\{d(Q_i), \sqrt{n\alpha}\}}{\tau}\right)$.
	By Claim~\ref{clm:snbr}, since \snbr\ is invoked with $\beta'$, its complexity is
	$O\left(\frac{\min\{d(Q_i),\sqrt{n\alpha\gamma}\}}{\tau}\cdot \log(1/\beta')\log n \right)$.
	Also observe that $d(Q_i)=d(C)$ (as noted in Observarion~\ref{obs:dQ}, if a $k$-clique $C$ is assigned to a $(k-1)$-clique $Q$, then $d(C)=d(Q)$).
	Therefore,
	the cost of the $i\th$ iteration of the loop in Step~\ref{step:se_j} is
	$O^*\left(\left \lceil \frac{\min\{d(C), \sqrt{n\alpha\gamma}\}}{\tau}\right\rceil \right )$.
	Furthermore, by the proof of  Lemma~\ref{lem:rw},	for every edge   $(u,u')\in E_F$,
	\[\Pr[(u,u')=(v_{Q_i},v_{Q_{i+1}})]\leq \frac{1+\beta}{X (s_k+1)}.\]
	Hence, the expected cost of the  $i\th$ invocation of the loop is
		\begin{align*} 
	\EX[cost]&=
	O^*\left(\sum_{(v_Q, v_{Q'})\in E_F} \;P_i[(v_Q, v_{Q'})]\cdot \left\lceil\frac{\min\{d(Q'), \sqrt{n\alpha\gamma}\} }{\tau}\right\rceil  \right)
	\\&=O^*\left(
	\frac{1}{X (s+1)\cdot \tau} \left( \sum_{C \in \mC_k \mid d(C)\leq \tau} 1 + \sum_{C \in \mC_k} d(C) \right)\right)\\& 
	=O^*\left(\frac{n\alpha^k}{X\tau}\right)
	=O^*(1)
	\end{align*}
	where the last inequality is by the assumptions that $X\cdot \tau=\Omega^*(n\alpha^{k})$ and  since $s=O^*(1)$.
	Hence, together with the recursive call and since there are $s$ invocations of the loop, the expected query complexity of the procedure is 
	\[O^*(q^{exp}_{n,\tau,\gamma}(k-1,\beta/(2s+2))+1).\]
	The maximum query complexity of the procedure is the maximum cost of the recursive call plus 
	the maximum cost of the invocations of the subroutines \snbr\ and \ISLZ\ which is bounded by
	$O^*\left(\frac{\sqrt{n\alpha}}{\tau} \right)$.
	Hence, the maximum cost is 
	\[
	O^*\left(q^{max}_{n,\tau,\gamma}(k-1,\beta/(2s+2))+ \frac{\sqrt{n\alpha}}{\tau} \right)
		.\]
\end{proof}

We now prove our main lemma, Lemma~\ref{lem:main_lem}, from which Theorem~\ref{thm:ub-with-tau}  and Corollary~\ref{cor:ub} follow.
\begin{lem}[Lemma~\ref{lem:main_lem}, restated]\label{lem:main_rec}
	Consider an invocation of \se$(H_k, \overline{N}_{k},\beta ,\vec{p}=(G,n,k,\tau,\alpha))$ where $\overline{N}_{k}\geq|V_{H_k}|,\frac{\alpha}{\beta}\cdot (4k\log n)^k< \tau<\sqrt{n\alpha}$.
	Let $\beta_i$ denote the value of $\beta$ with which the recursive call \se\ for $H_i$ was invoked (so that in particular $\beta_k=\beta$).
	
	The procedure returns every edge in $H_k$ with probability in \begin{equation*}\frac{1\pm\beta_2}{n\alpha\gamma\cdot \tau^{k-2}}\cdot  \prod_{i=3}^k\frac{ \beta_i^2\cdot (1\pm\beta_i)}{(40i(s_i+1)\cdot \log^2(n/\beta_i))}.
	\end{equation*}
	
	Furthermore, the expected query complexity of the procedure is
	$O^*(1)$,
	and the maximum query complexity is
	$O^*\left( \frac{\sqrt{n\alpha}}{\tau}\right)$.
\end{lem}
\begin{proof}
	We prove the lemma by induction on $k$.
	If invoked with $F=H_2$ and $\beta_2$ then $F=G$, and
	By Theorem~\ref{thm:ERR}, each oriented edge in $H_2$ is returned with probability $\frac{1\pm\beta_2}{n\alpha\gamma}$ for $\gamma=\setGamma$. Therefore, each edge is returned with probability $\frac{2(1\pm\beta_2)}{n\alpha\gamma}$.
	
	We  now assume that the induction claim holds for $k-1 \geq 2$ and prove it for $k$.
	That is, we assume that  if \se\ is invoked
	with parameters $H_{k-1}\leq  \overline{N}_{k-1}, \tau$ and $\beta_{k-1}$, 
	such that $4\alpha/\beta_{k-1}\leq \tau \leq \sqrt{n\alpha}$, then 
	each $(k-1)$-clique in $H_{k-1}$ is sampled with probability  $\frac{1\pm\beta_2}{n\alpha\gamma\cdot \tau^{k-3}}\cdot  2\prod_{i=3}^{k-1}\frac{ \beta_i^2\cdot (1\pm\beta_i)}{(40i(s_i+1)\cdot \log(\oN_i/\beta_i))}$. 
	
	We shall prove that  \se$(H_{k}, \overline{N}_{k},\beta, \vec{p}=(G,n,k,\tau, \alpha))$ returns each $k$-clique with probability $\frac{1\pm\beta_2}{n\alpha\gamma\cdot \tau^{k-2}}\cdot  2\prod_{i=3}^k\frac{ \beta_i^2 \cdot (1\pm\beta_i)}{(40i(s_i+1)\cdot \log(\oN_i/\beta_i))}$.
	By Lemma~\ref{lem:rw}, it is sufficient to prove that
	\begin{compactenum}
		\item 	If \ISLZ\ is $\tau$-good with respect to $H_{k}$ then 	
		\slz$(F,\oN_{F}, \beta', \tau,\vec{p})$ invoked in Step~\ref{step:sallow} is a $(\beta',X)$-good $E(L_0)$-sampling subroutine  for $L_0$ that is determined by \ISLZ.
		\item The subroutine \ISLZ$(H_k,\oN_k, v,\delta, \beta',\tau, \vec{p})$  is a $(\delta, \tau)$-good $L_0$-oracle with respect to $H_k$ for $\delta'\leq \beta'/X$.
		\item The subroutine \snbr$(F, v, \beta', \tau,\vec{p})$ invoked in Step~\ref{step:samp_nbr} is a $\beta'$-good neighbor-sampling subroutine for $F$.
\end{compactenum}

	For the first item, we would like to prove that the conditions of Claim~\ref{clm:slz} regarding \slz\ are met, building on the recursion hypothesis. 
	First observe that for every $i$, $\beta_{i-1}=\beta_i/(2s_i+2)$ for $s_i=\lceil \oN_i\rceil=O(k\log n)$.
	Hence, by the assumption  $\frac{4\alpha}{\beta}(4k\log n)^k \leq \tau$ ,for every $i$, 	
	$\beta_i \cdot \tau>4\alpha$, and in particular for $i=k-1$.  
	Also, by Claim~\ref{clm:bound-clqs}, for every $i$, $|H_i|\leq  n\alpha^{i-1}=\oN_i$, so that by Corollary~\ref{cor:ISLZ_oracle}, \ISLZ\ is a $(\delta,\tau)$-good oracle with respect to $H_{k-1}$.
	Hence, the
	recursive invocation of \se\ in Step~\ref{step:slz_rec} meets the assumptions in the current lemma, and therefore, by the induction hypothesis, it holds that \se\ as invoked in Step~\ref{step:slz_rec} is a $(\beta_{k-1},X_{k-1})$-good $(k-1)$-clique sampling procedure for $X_{k-1}$ such that
	each $k$-clique is returned with probability \[\frac{1\pm\beta_2}{n\alpha\gamma\cdot \tau^{k-3}}\cdot   2\prod_{i=3}^{k-1}\frac{ \beta_i^2 \cdot (1\pm\beta_i)}{(40i(s_i+1)\cdot \log^2(n/\beta_i))} = \Omega^*\left(\frac{1}{n\alpha\tau^{k-2}} \right)\;.
	\]
	Since $n\alpha^{k-2}<n\alpha\tau^{k-3}$, it holds that $X_{k-1}\tau^2=\Omega^*(n\alpha^{k}).$
	Therefore, in the case that $L_0$ is $\tau$-good then the conditions of Claim~\ref{clm:slz} hold, implying that the first item holds.
	The second item holds by Corollary~\ref{cor:ISLZ_oracle}, and the third item holds by Claim~\ref{clm:snbr}.
	
	Hence, by Lemma~\ref{lem:rw}, the procedure \se\ returns every $k$-clique in $G$ with probability
	$\frac{1\pm\beta}{X'}$ for $X'=X \cdot\tau(s_k+1)\cdot 40k\log^2(n/\beta_k)$.
	Hence, each $k$-clique in $H_k$
	is sampled with probability
	\begin{align*}
	&\frac{1\pm\beta_2}{n\alpha\gamma\cdot \tau^{k-3}}\cdot   2\prod_{i=3}^{k-1}\frac{ \beta_i^2 \cdot (1\pm\beta_i)}{(40i(s_i+1)\cdot \log^2(n/\beta_i))}\cdot \frac{\beta_k(1\pm\beta_k)}{40k(s_k+1)\cdot \log^2(n/\beta_k)\cdot \tau}
	\\=&
	\frac{1\pm\beta_2}{n\alpha\gamma\cdot \tau^{k-2}}\cdot   2\prod_{i=3}^k\frac{ \beta_i^2 \cdot (1\pm\beta_i)}{(40i(s_i+1)\cdot \log^2(n/\beta_i))}
	.\end{align*}
	This concludes the first item of the claim.

	We now turn to analyze the expected and maximum query and time complexities of the procedure. Let $q_k$ denote the expected query complexity of \se\ when invoked on the graph $H_k$.
	Here too we prove the claim by induction on $k$, starting from $k=2$. By Theorem~\ref{thm:ERR}, \SBE\ has    $O(\log n)$ query and time complexity, so that $q_2=O(\log n)$ and the claim holds.
	We assume the claim holds for $k-1$ and prove it for $k$.
	By Claim~\ref{clm:slz}, each invocation of \slz\ has expected cost $O^*(q_{k-1}+1)$. Therefore, by Lemma~\ref{clm:se_comp}, the expected query complexity of \se\ when invoked for $H_k$ is
	$O^*\left(q_1 \right)$.
	Therefore,
	the query complexity and running time are bounded by	$O^*\left(q_1 \right)$.
	By Claim~\ref{clm:se_comp}, the maximum query and time complexity of the $O(k)$ recursive invocations is
	$O^*\left( \log n+\frac{\sqrt{n\alpha}}{\tau}\cdot k \right)=
	O^*\left( \frac{\sqrt{n\alpha}}{\tau}\right).
	$
	This concludes the proof of the lemma.
\end{proof}

We next prove Theorem~\ref{thm:ub-with-tau}.
\begin{proof}[Proof of Theorem~\ref{thm:ub-with-tau}]
	Consider an invocation  \sclq$(G,k,\alpha, \eps, \onk, \tau)$ with $\onk\in [n_k, 2n_k]$ and $\overline{N}_{k}\geq|V_{H_k}|,\frac{\alpha}{\eps}\cdot (4k\log n)^k< \tau<\sqrt{n\alpha\gamma}$ for $\gamma=\setGamma$.
	
	By Lemma~\ref{lem:main_rec}, each invocation of \se\ with $H_k$ returns a each $k$-clique with probability
	$\frac{1\pm\beta_2}{n\alpha\gamma\cdot \tau^{k-2}}\cdot  \prod_{i=3}^k\frac{ \beta_i^2 \cdot (1\pm\beta_i)}{(40i(s_i+1)\cdot \log^2(n/\beta_i))}.$
	
	Since for every $i$, $\beta_{i-1}=\beta_{i}/(2s_i+2)$ and $s_{i}=\log (n\alpha^{i})$, it holds that $\beta_i =\Omega(\beta_k/(k^2 \log n)^i)$.
	Hence,
	by the setting of $\beta_k=\eps/10k$,
	the overall success probability of outputting any clique in a single invocation is
	\[
	\Omega\left(
	n_k \cdot \frac{(1-\beta_k)^k}{n\alpha\gamma\cdot \tau^{k-2}} \cdot \left(\frac{\beta_k}{k\log n}\right)^{o(k)}
	\right)= \frac{n_k}{n\alpha\gamma\cdot \tau^{k-2}} \cdot \Omega\left(
	 (\beta/(k\log n))^{O(k)}
	\right).
	\]

	Hence, the expected number of invocations until a $k$-clique is returned is $O\left( \frac{n\alpha\tau^{k-2}}{\tau}\right) \cdot (4k\log n/\eps)^{O(k)}$.
	By Lemma~\ref{lem:main_rec}, the expected cost of each invocation is $O^*(1)$, so that it follows that the expected complexity of the algorithm is 
	\[ O^*\left(\frac{n\alpha\tau ^{k-2}}{n_k}\right).	\]
	Therefore the claim regarding the expected complexity holds.
	
	It also follows that there exists a sufficiently large constant $c'_1$, such that if we perform at least $t'_1=\frac{n\alpha\tau^{k-2}}{n_k}\cdot (k\log n /\eps)^{c'_1 \cdot k}$ invocations, then with probability at least $1-\eps/10n_k$, a $k$-clique is returned. 
	Hence, if we allow $t''_1=\frac{n\alpha\tau^{k-2}}{n_k}\cdot (k\log n /\eps)^{c_1 \cdot k}$ queries (in $G$) for a sufficiently large $c_1>c'_1$, then with probability at least $1-\eps/10n_k$, a $k$-clique is returned. 
	We note that if $t''_1$ exceeds $n\alpha$, then the algorithm may simply read the entire graph by querying the neighbors of all vertices, which takes at most $n\alpha$ queries. Hence,  we can let $t_1=\min\{t''_1, n\alpha\}$.

	The claim regarding the maximum query complexity holds by the setting of $r$ in Step~\ref{step:sclq_halt} of the algorithm. Hence, it remains to prove that the resulting distribution on $k$-cliques is pointwise $\eps$-close to uniform.
We start by showing that with probability at least $1-\eps/10n_k$, the 	algorithm does not halt in Step~\ref{step:sclq_halt}.
	
Consider random variables $\chi_j$ such that $\chi_j$ is the running time of the $j\th$ invocation.
By Lemma~\ref{lem:main_rec}, for every $j$,
$\EX[\chi_j]=O^*(1)$, and  $\max\{\chi_j \}=O^*(\sqrt{n\alpha}/\tau)$.
Hence, by the multiplicative Chernoff bound, if we perform at least $t_2=\frac{\sqrt{n\alpha}}{\tau}\cdot (k\log n/\eps)^{c_2\cdot k}$ invocations for a sufficiently large constant $c_2$, then we have that, with probability at least $1-\eps/10n_k$, the number of queries does not exceed its expected value $O^*(t_2)=O^*(\sqrt{n\alpha}/\tau)$.
Therefore, if we halt the algorithm after performing 
at most $t_3=\max\left\{\min\left\{n\alpha,\frac{n\alpha\tau^{k-2}}{n_k}\right\},\frac{\sqrt{n\alpha}}{\tau} \right\}\cdot (k\log n/\eps)^{c\cdot k}$ queries for a suffiecently large constant $c$, it holds that with probability at least $1-\eps/10n_k$, a $k$-clique is returned.
Hence, with probability at least $1-\eps/5n_k$, the algorithm does not fail in Step~\ref{step:sclq_halt} (since $t_3\geq t_2$) and it returns a $k$-clique (since $t_3\geq t_1$).	

	Finally, conditioned on a $k$-clique being returned, each $C$ is returned with probability in $\frac{1\pm\eps/10}{(1\pm\eps/10)n_k}$.
	Hence, after $t$ invocations, each $k$-clique is returned with probability
	$\left[\frac{(1-\eps/5)(1-\eps/10)^2}{n_k}, \frac{1+\eps/10}{n_k}\right] \in \frac{1\pm \eps}{n_k}$
	so that the resulting distribution on $k$-cliques is $\eps$-close to uniform.
	This concludes the proof of the theorem.
\end{proof}

\begin{proof}[Proof of Corollary~\ref{cor:ub}]
	
	If the algorithm is not given the estimate $\onk$ and a parameter $\tau$ then it proceeds as follows.

	First, it invokes the $k$-clique counting algorithm by~\cite{ERS20_soda} with an approximation parameter $\eps=1/4$   for $t=\log(10n_k/\eps)$ times. It then sets $\onk$ to be twice the median of the returned estimates.  
	Since each invocation of the counting algorithm returns an estimate $\onk\in (1\pm1/4)n_k$ with probability at least $2/3$, it holds that with probability at least $1-\eps/5n_k$, $\onk\in[n_k, 2n_k]$.
	As the expected complexity of each such invocation is $O^*(n\alpha^{k-1}/n_k)$,
	these invocations do not asymptotically affect the complexity of the algorithm. However,  as the bound on the complexity of the counting algorithm is only on the expected value, and it is not guaranteed to hold with high probability, we must remove the condition of Step~\ref{step:sclq_halt}, so that our guarantee regarding the complexity of the algorithm only holds \emph{in expectation}.
		
	The algorithm then sets $\tau=\settau$. Clearly for both cases  $\tau\geq \frac{\alpha}{\eps}\cdot (4k\log n)^k$.
	Also, if $\tau=\frac{\alpha}{\eps}\cdot (4k\log n)^k$ then $\tau\leq \sqrt{n\alpha}$.
	Otherwise, for $\tau=\frac{\onk^{1/k}}{(n\alpha)^{1/(2(k-1))}}$,  since $n_k\leq(n\alpha)^{k/2}<(n\alpha)^{(k/2)(k/(k-1))}$, it  holds that $\tau<\sqrt{n\alpha}$.

 Finaly it sets $\eps'=\eps/3$ and runs the algorithm as before (with the new $\eps'$ value) and without the halting condition in Step~\ref{step:sclq_halt}.

	By the abvoe settings, if $\onk\in[n_k, 2n_k]$, then   the conditions of Theorem~\ref{thm:ub-with-tau} hold. Therefore, if $\onk \in [n_k , 2n_k]$, it holds that the resulting distribution on $k$-cliques is pointwise $(\eps/3)$-close to uniform. Since the condition holds with probability at least $1-\eps/5n_k$, it follows that the resulting distribution on $k$-cliques is pointwise $\eps$-close to uniform.	
\end{proof}

\newcommand{\ow}{\overline{w}}
\newcommand{\oW}{\overline{W}}
\newcommand{\oG}{\overline{G}}
\newcommand{\cnsta}{4}
\newcommand{\cnstb}{8}
\newcommand{\oS}{\overline{S}}

\newcommand{\tG}{\widetilde{G}}

\newcommand{\ncalW}{\overline{\mathcal{W}}}
\newcommand{\calW}{{\mathcal{W}}}

\newcommand{\dmaxx}{{\tilde{d}}}
\newcommand{\ovE}{{\overline{E}}}
\newcommand{\ovGamma}{{\overline{\Gamma}}}

\newcommand{\hG}{{\widehat{G}}}

\newcommand{\kG}[1]{{G_{#1}^{\rm kn}}}
\newcommand{\kd}[1]{{d_{#1}^{\rm kn}}}
\newcommand{\mW}{\mathcal{W}}

\renewcommand{\cA}{\textsc{Alg}}
\newcommand{\mQ}{\mathcal{Q}}
\newcommand{\emphsf}[1]{\textsf{#1}}

\section{A Lower Bound for Sampling Cliques}\label{sec:lb-tri}
In this section we prove a lower bound on sampling $k$-cliques from a distribution that is $\eps$-pointwise close to uniform.
We show that $\Omega\left(\max\left\{\left(\frac{(n\alpha)^{k/2}}{k^k\cdot n_k}\right)^{\frac{1}{k-1}},\frac{n\alpha^{k-1}}{n_k}\right\}\right)$ queries are necessary in order to ensure that
each $k$-clique is sampled with probability at least $1/(2n_k)$.

The second term in this lower bound follows from a related lower bound on approximate counting of the number of $k$-cliques.  We discuss this term briefly in Section~\ref{subsec:lb-counting}.  The main focus of this section is on the first term.  Observe that as long as $n_k \leq k^{\frac{k}{k-2}}n^{\frac{1}{2}}\alpha^{k-\frac{1}{2}}$, the second term is dominant. We hence establish the lower bound of $\Omega(Q)$ for $Q =\left(\frac{(n\alpha)^{k/2}}{k^k\cdot n_k}\right)^{\frac{1}{k-1}}$, assuming that $n_k > k^{\frac{k}{k-2}}n^{\frac{1}{2}}\alpha^{k-\frac{1}{2}}$, so that $Q$ is smaller than $\sqrt{n/\alpha} < \sqrt{n\alpha}$ by a sufficiently large constant factor. 

In a nutshell, the lower bound of $\Omega(Q)$  is based on ``hiding'' a $k$-clique and proving that any algorithm that outputs this clique with probability $\Omega(1/n_k)$, must perform $\Omega(Q)$ queries. This complexity (almost) matches the upper bound of Theorem~\ref{thm:ub_informal}, thus proving the optimality of our algorithm (up to a $(\log n/\eps)^{O(k)}$ factor).

\subsection{The lower bound construction}
We define a family of graphs, denoted
$\cG$, where all graphs in the family have the same basic structure, and only differ in the labeling of the
vertices and edges. The graphs are over $n$ vertices and have arboricity at most $\alpha$.
For a given integer $\tnk$, all graphs contain $n_k = \Theta(\tnk)$ $k$-cliques.

In what follows we assume for simplicity (and without loss of generality), that various expressions are integers (e.g., $\sqrt{n\alpha}$). We also assume that $n$ is sufficiently large.The basic underlying graph structure, denoted $\tG = (V,E)$, is as follows.
The set of vertices, $V$  is partitioned into four subsets: $A$, $B$, $C$ and $D$.
For $n'$ that satisfies $n= 2n' + \sqrt{n'\alpha}\cdot k$, we have that $|A|=\ell = \sqrt{n'\alpha}$, $|B|=n'$, $|C|=\ell\cdot (k-1)$ and $|D|=n'$. Furthermore, the vertices in $A$ have labels in $[\ell] = \{1,\dots,\ell\}$, the vertices in $B$ have labels in $\{\ell+1,\dots,\ell+n'\}$, the vertices in $C$ have labels in $\{\ell+n'+1,\dots,\ell\cdot k+n'\}$ and the vertices in $D$ have the remaining labels.
Each edge $\{u,v\}$ corresponds to two oriented edges, $(u,v)$ and $(v,u)$, where each has its own label (the label of $(u,v)$ is in $[d(u)]$ and the label of $(v,u)$ is in $[d(v)]$).

\begin{figure}[h]
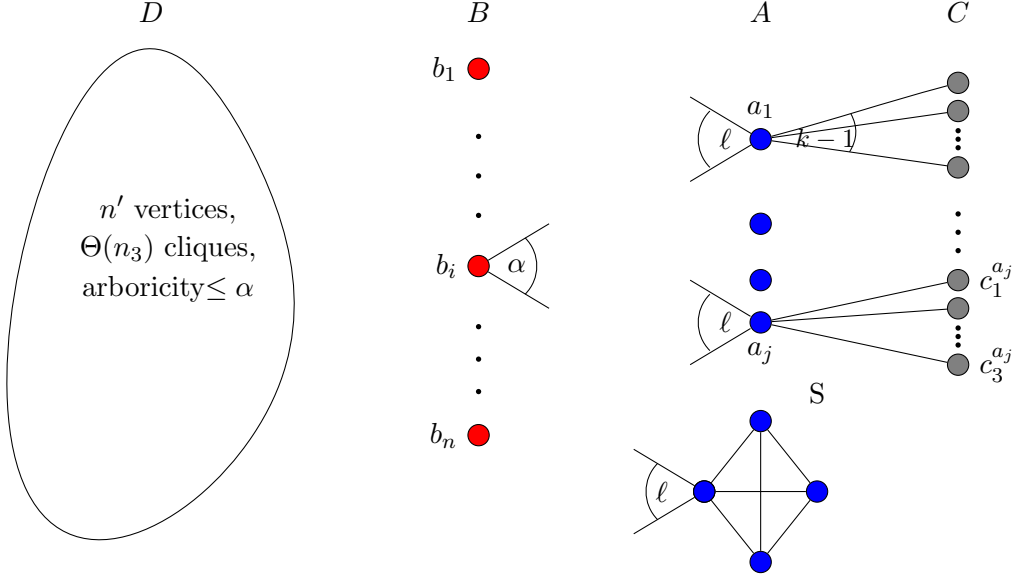
\label{fig:LB}
	\centering
	\drawLB
	\caption{The underlying structure of the graphs in $\cG$ for $k=4$.}
\end{figure}

The subgraph of $\tG$ induced by $D$ is a fixed subgraph that contains $n_k-1 = \Theta(\tnk)$ $k$-cliques (and has arboricity at most $\alpha$).
Each vertex in $A$ has $\ell$ neighbors in $B$, and each vertex in $B$ has $\alpha$ neighbors in $A$ (observe that indeed $|A|\cdot \ell = \ell^2 = n'\alpha$ and $|B|\cdot \alpha = n'\alpha$). This bipartite subgraph is fixed, and furthermore, the label of each (oriented) edge $(v,u)$ for $v\in B$ and $u\in A$ is fixed as well.
However, the labels of the edges $(u,v)$ (for $u\in A$, $v\in B$) are not fixed, and vary between graphs in $\cG$.
The graphs in $\cG$ also differ in the choice of a subset $S\subset A$ of $k$ \emph{special vertices}, where the subgraph induces by $S$ is a $k$-clique. Finally, each vertex $u \in A\setminus S$ has $k-1$ distinct neighbors in $C$
(where the choice of these neighbors differs as well between graphs in $\cG$).
Observe that all vertices in $A$ have  degree $\dmaxx=\ell+(k-1)$, all vertices in $B$ have degree $\alpha$ and all vertices in $C$ have degree $1$. See Figure~\ref{fig:LB} for an illustration of a graph in $\cG$ for the case of $k=4$.
As noted above, while all graphs in $\cG$ have the same underlying structure, as labeled graphs they differ in the subgraph induced by $A\cup C$ (and in particular in the  choice of the special subset $S \subset A$), as well as in the labels of the edges $(u,v)$ for $u\in A$ and $v\in B$.

\subsection{The process $\calP$ and the knowledge graph}
We now define a process $\calP$ that answers queries of a sampling algorithm $\cA$ while constructing a uniformly selected graph $\hG \in \cG$ on the fly.  We shall be interested in showing that
 if $\cA$ performs less than $Q/c$ queries, for a sufficiently large constant $c$,
 then the probability that it outputs the labels of the vertices in the set $S$ is less than $1/(2 n_k)$.
 Given our assumption that the labels of these vertices and their neighbors belong to $[n'+\ell\cdot k]$, we may assume without loss of generality that $\cA$ does not perform queries with labels in $\{n'+\ell \cdot k,\dots,2n'+\ell\cdot k\}$ (i.e., labels of vertices in $D$).

It will be helpful to consider what
we refer to as the \emph{knowledge graph\/} of $\cA$, after it performed $t$ queries, denoted $\kG{t} = (V_t,E_t,\ovE_t,\tau_t,d_t)$.
Here $V_t$ is the set of (labeled) vertices in the knowledge graph,
 $E_t$ is the set of (oriented) edges, 
$\ovE_t$ is the set of ``non-edges'' (corresponding to negative answers to pair queries),
$\tau_t$ is the edge-labeling function of $\kG{t}$, and $d_t$ is the degree function of $\kG{t}$.
More precisely, for each vertex $u$ that appeared either in one of the first $t$ queries  of $\cA$ or was an answer to one of these queries, there is a vertex in $V_t$. If for $t' \leq t$  a neighbor query $q_{t'} = \nbr(u,i)$ was answered by $v$, then there is an edge $(u,v) \in E_t$ labeled by $i$ and $\tau((u,v))=i$.
We shall assume that $\calP$ also provides $\cA$ with the label $j$ of the edge $(v,u)$,
so that $(v,u) \in E_t$ as well, and $\tau_t((v,u))=j$.
If for $t' \leq t$ a pair query $q_{t'} = \pair(u,v)$ 
was answered positively, then
$(u,v), (v,u) \in E_t$, and if it was answered negatively, then $(u,v), (v,u) \in \ovE_t$.
 In the former case we assume that $\cA$ is provided with the labels of the edges $(u,v)$ and $(v,u)$, so that $\tau_t((u,v))$  and $\tau_t((v,u))$ are set appropriately. Finally, for each $u \in V_t$, $d_t(u)$ is the degree of $u$ in $\kG{t}$. Note that $d_t(u) \leq d(u)$ where $d(u)$ is the degree of the vertex $u$ in the final graph $\hG$ (e.g., $d(u)=\dmaxx$ for each $u\in A$).

For $X \in \{A,B,C\}$ let
 $V_{t,X} \subseteq V_t$ denote the subset of
 vertices $u$ in $\kG{t}$ that belong to $X$.
  For each $u \in V_t$, let $\Gamma_t(u)$ denote its set of neighbors in $\kG{t}$,
  and for $X \in \{A,B,C\}$ let $\Gamma_{t,X}(u)=\Gamma_t(u)\cap V_{t,X}.$
 We use $\ovGamma_t(u)$ to denote the subset of ``non-edges'' $(u,v) \in \ovE_t$  
 and $\overline{d}_t(u)=|\ovGamma_t(u)|$.
  Finally, let $S_t \subseteq A$ denote the subset of
  vertices $u \in A$ such that
  $|\Gamma_{t,A}(u)| > 0$,
  and let $\oS_t$ 
  denote the subset of
  vertices $u \in A$ such that
  $|\Gamma_{t,C}(u)|>0$. Hence, if $u\in S_t$, then it is one of the $k$ vertices in the
  hidden $k$-clique, and if $u\in \oS_t$, then it does not participate in any $k$-clique (so $u\in A\setminus S$ for the special subset $S$).

 We refer to the edges 
  between $A$ and $C$ as \emph{informative} edges (since they indicate whether a vertex  in $A$ belongs  to the special set $S$ or to $A\setminus S$), and to the edges 
 between two vertices of $A$ as \emph{witness} edges. 

\subsection{Details on how $\calP$ answers queries}
 The process $\calP$ answers  the queries of $\cA$  as follows. To answer a query $q_{t}$, $\cP$ considers the subset of (labeled) graphs in $\cG$ that are consistent with the knowledge graph $\kG{t-1}$, which we denote by $\cG_{t-1}$. It then selects a graph $\hG_t \in \cG_{t-1}$ uniformly at random and answers the query according to $\hG_t$. Finally, it updates $\kG{t-1}$, to obtain $\kG{t}$ by incorporating the new information obtained from the answer to this query (including the additional information as described previously). 
 Once all $T$ queries are performed, $\calP$ uniformly selects a graph $\hG$ in $\cG_T$ and this is the resulting graph. Note that the graphs $\hG_t$ for $t < T$ are only used as a tool 
 to aid the description of the process, so that each $\hG_t$ can be viewed as being ``discarded'' after answering $q_t$.
Observe that this  process generates a uniformly distributed graph in $\cG$.

We shall say that $\cA$ \emph{succeeds\/}, if,
after performing $T$ queries, it outputs a subset $\widehat{S}$, such that $\widehat{S}$ equals the special subset $S$ (the hidden clique) of the final graph $\hG$.
In particular this is the case if $|S_T| = k$. However, the algorithm may output the special subset even if $|S_T|<k$.
To address this possibility, it will be convenient to assume (without loss of generality), that in order to select the final graph $\hG$, the process first determines whether each of the vertices in $\widehat{S} \setminus S_T$ belongs to the hidden clique (in the same manner that it answers queries of the algorithm).

In order to analyze the success probability of any algorithm $\cA$, we introduce the following central notions.
 \begin{dfn}[Witness answers]\label{def:witness}
  We say that an answer $a_t$ to a query $q_t$ (given the knowledge graph $\kG{t-1}$) is an \emphsf{edge-witness} answer, if two new vertices of the hidden clique are discovered. Namely, $|S_t\setminus S_{t-1}|=2$.
  This event is denoted by 
  $\calE^2_t$.
  Similarly, we say that an answer $a_t$ is a \emphsf{vertex-witness} answer if one  such new vertex is discovered, that is, $|S_t\setminus S_{t-1}|=1$.
  This event is denoted by $\calE^1_t$. 
 \end{dfn}
Observe that if $\calE^1_t$ occurs then necessarily the event $\calE^2_{t'}$ occurred for some $t'<t$. 	

We shall prove  that the following holds for any algorithm $\cA$ that performs
 $T \leq \ell/\cnsta$ queries: 
 (1)
 each answer has probability 
 at most $\cnstb k^2/\ell^2$ 
 to be an edge-witness answer, and (2) following the first edge-witness answer (the first edge in the hidden $k$-clique),  each answer has probability at most
 $\cnstb k/\ell$
 to be a vertex-witness answer.
 We may assume without loss of generality that $\cA$ does not perform pair queries that include a vertex $v\in C$,
 so that $\ovE_t$ never contains such a pair. This is the case since such a pair query can be replaced by a neighbor query $(v,1)$, which returns the specific single neighbor $u \in A$ of $v$, and as a consequence also determines that there is no edge between $v$ and any other vertex in $A$.

\begin{figure}[h]
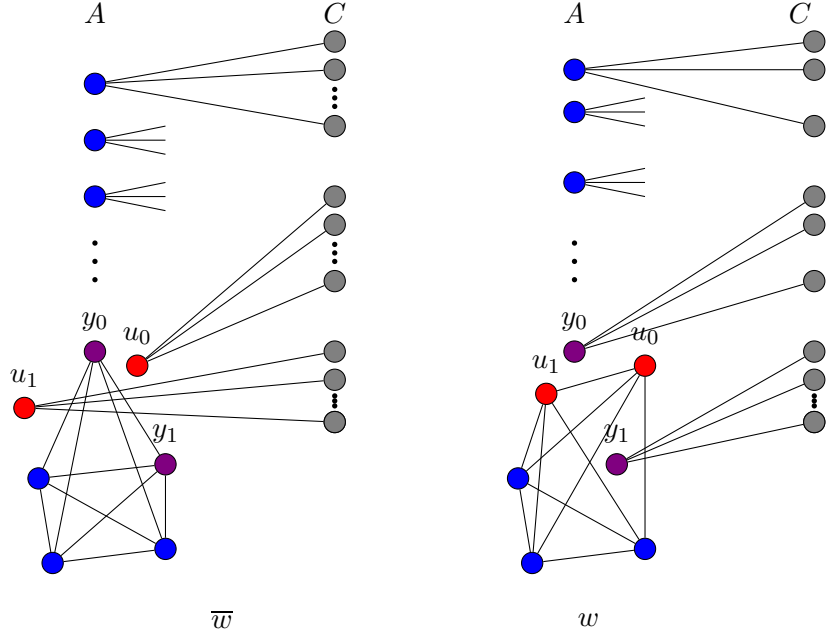
 \label{fig:switch}
	\centering
	\drawSwitch
	\caption{Two neighboring graphs in $H_t$ for $k=5$. }
\end{figure}

\begin{clm}\label{clm:witness-edge}
	For every $t\leq \ell/\cnsta$, 
       every knowledge graph $\kG{t-1}$ and every query $q_t$, 
	\[  
     \Pr[\calE^2_{t}]
             \leq \frac{\cnstb k^2}{\ell^2}\;.
	\]
\end{clm}
\begin{proof}
	First consider the case that $q_{t} = \pair(u,u')$ for some
$u,u'\in A$. Note that unless $u,u'\in A \setminus (S_{t-1}\cup \oS_{t-1})$ (and $(u,u') \notin \ovE_{t-1}$), we have that $\Pr[\calE^2_{t}]=0$. This holds since $u,u'\notin S_{t-1}$ by the definition of $\calE^2_t$, and
if either $u$ or $u'$ belongs to $\oS_{t-1}$, then clearly the response cannot correspond to a witness edge.

	
	In order to prove the claim (for pair queries), we shall bound the fraction of graphs in $\cG_{t-1}$ such that $u$ and $u'$ belong  to the hidden clique.
Refer to such graphs as \emph{witness} graphs,  and to all other graphs in $\cG_{t-1}$ as \emph{non-witness} graphs. We denote the set of witness graphs (with the witness pair $(u,u')$) by $\calW_t = \calW_t(u,u')$, and the set of non-witness graphs by $\ncalW_t = \ncalW_t(u,u')$. (Here we use the subscript $t$, since while these graphs belong to $\cG_{t-1}$, they also depend on $q_t=\pair(u,u')$.)
	In order to bound the fraction of witness graphs in $\cG_{t-1}$, we define the following auxiliary graph $H_{t}$.
	The graph $H_{t}$ is a bipartite graph over the sets of nodes $W$ and $\oW$, where  in $W$ there is a node for every
graph in $\calW_t$
and in $\oW$ there is a node for every 
graph in $\ncalW_t$.


Let $\hG$ be a witness graph in $\calW_t$ with the special set
 $S=\{u_0=u,u_1=u',r_1, \ldots, r_{k-2}\}$, and let $w$ be the node it corresponds to in $H_t$.
 For each  two vertices $y_0,y_1\in A\setminus (S \cup \oS_{t-1})$ such that
 $(y_0,y_1)\notin \ovE_{t-1}$ and
  $(y_b,r_p) \notin \ovE_{t-1}$ for every $b\in \{0,1\}$ and $p\in [k-2]$, there is a neighbor $\bar{w}$ of $w$ in $H_t$, corresponding to the graph $\hG'\in \ncalW_t$ that results from $\hG$ by performing the following operations.

For $b \in \{0,1\}$,
let $v_{b,1},\dots,v_{b,k-1}$ be the neighbors of $y_b$ in $C$.
First, we remove the edge $(u_0,u_1)$ and add the edge $(y_0,y_1)$. 
Next, for each $b\in \{0,1\}$ we remove the edge $(u_b,r_p)$ for each
$p\in[k-2]$, and the edge $(y_b,v_{b,p})$ for each $p\in [k-1]$.
Finally, for each $b\in \{0,1\}$, we
add the edges $(y_b,r_p)$ for each $p \in [k-2]$ and the edges $(u_b,v_{b,p})$ for each $p \in [k-1]$.
(Recall that we assumed without loss of generality that there are no pairs in $\ovE_{t-1}$ that contain a vertex in $C$, so that such an addition is consistent with $\kG{t-1}$.) See Figure~\ref{fig:switch} for an illustration.

It remains to specify the labels of all new  edges.
The  new (oriented) edge $(y_0,y_1)$ is given the label of the removed edge $(y_0,v_{0,k-1})$, and $(y_1,y_0)$ is given the label of $(y_1,v_{1,k-1})$.
For $b\in \{0,1\}$, each new edge $(y_b,r_p)$ for $p\in [k-2]$ is given the label of the removed edge $(y_b,v_{b,p})$,
each new edge $(u_b,v_{b,p})$ for $p \in [k-2]$ is given the label of the removed edge $(u_b,r_p)$, and
$(u_b,v_{b,k-1})$ is given the label of $(u_b,u_{1-b})$.
	
By the description of the process $\cP$, for any pair $(u,u')$ (where $u,u'\in A \setminus (S_{t-1}\cup \oS_{t-1})$ (and $(u,u') \notin \ovE_{t-1}$)
\[
\Pr[\calE^2_t\;|\; q_t = \pair(u,u')]=\frac{|\calW_t|}{|\cG_{t-1}|}=\frac{|W|}{|W\cup \oW|}.
\]
Let $d_W$ and $d_{\oW}$ denote the average degree of the nodes in $W$ and $\oW$, respectively.
Since $H_t$ is a bipartite graph, $|W| =\frac{|\oW|\cdot d_{\oW}}{d_W}$. Hence, to bound $\Pr[\calE^2_t]$, we will be interested in bounding $|W|$ by lower bounding $d_W$ and upper bounding $d_{\oW}$.

By the definition of the set of neighbors of each node $w$ in $W$,
the number of ``eligible'' pairs of vertices $(y_0,y_1)$ that can be used to define a neighbor $\bar{w}$ of $w$
is lower bounded by $\binom{|A| - k - |\oS_{t-1}| - |\ovE_{t-1}|}{2}$.
Since 
$|\oS_{t-1}| + |\ovE_{t-1}| \leq  \ell/\cnsta$,
there are at least $\ell^2/\cnstb$ such pairs.
Therefore, $d_W \geq \ell^2/\cnstb$. 

Now consider a non-witness graph $\hG' \in \ncalW_t$.
Since $\hG'$ is a non-witness graph,  $u$ and $u'$ do not belong to the special set  $S' = \{r_1, \ldots, r_{k}\}$ of $\hG'$. By the definition of $H_t$, each neighbor in $H_t$ of the node $\bar{w}_{\hG'}$ corresponding to $\hG'$ must correspond to a graph $\hG \in \calW_t$ that contains a $k$-clique over $u,u'$ and $k-2$ of the vertices in $S'$. Therefore, $\bar{w}_{\hG'}$ has at most ${k\choose 2}$ neighbors in $H_t$. Since this is true for every node $\bar{w} \in \oW$, we have that $d_{\oW} < k^2$.
Therefore,  $|W| \leq \frac{|\oW|\cdot k^2}{\ell^2/\cnstb})$, and so
\[
\Pr[\calE^2_t \;|\; q_t = \pair(u,u')]=\frac{|W|}{|W\cup \oW|}\leq \frac{|W|}{|\oW|}\leq \frac{\cnstb k^2}{\ell^2}\;.
\]

The analysis of the case that $q_t$ is a neighbor query  $\nbr(u,i)$ is essentially the same.
If $a_t=u'$ is the returned neighbor, then, here too, $(u,u')$ may be is a witness edge only if
$u,u'\in A\setminus (S_{t-1}\cup \oS_{t-1})$ (and $(u,u') \notin \ovE_{t-1}$). The set of witness and non-witness graphs is defined for each possible answer $u'$, where in each witness graph, the label of $(u,u')$ is $i$.
By the same reasoning as above (for every $u$, $i$ and $u'$),
$\Pr[\calE^2_t \;|\; q_t =\nbr(u,i) \;\&\; a_t=u'] \leq \cnstb k^2/\ell^2$, and the claim is established.
\end{proof}

 \begin{clm}\label{clm:witness-vertex}
 	For every $t\leq \ell/\cnsta$, 
 every knowledge graph $\kG{t-1}$ and every query $q_t$, 
 	\[ \Pr[\calE^1_t] \leq \frac{\cnstb k}{\ell}\;.
 	\]
 \end{clm}
 \begin{proof}
 The proof of this claim is similar to the proof  of Claim~\ref{clm:witness-edge}, with some small changes to the definition of the auxiliary graph $H_t$. 
 We analyze the case that $q_t$ is a pair query, and the case the $q_t$ is a neighbor query is essentially the same.
 If $q_{t} = \pair(u,u')$, then $\calE^1_t$ may occur only if
 $u\in S_{t-1}$ and $u'\in A\setminus (S_{t-1}\cup \oS_{t-1})$, or vice versa
 (and $(u,u') \notin \ovE_{t-1}$).

 	Here too we define an auxiliary bipartite graph $H_t$ over a set of nodes $W\cup \oW$, where there is a node in $W$ for each graph
 in $\calW_t$, and a node in $\oW$ for each graph in $\ncalW_t$.
 There is an edge between a node in $W$ and a node in $\oW$ only if the corresponding graphs differ on a \emph{single} vertex in the hidden clique (rather than two vertices, as was the case in the proof of Claim~\ref{clm:witness-edge}).
 More precise details follow.
 	

Let $\hG$ be a witness graph in $\calW_t$ with the special set
 $S=\{u,u',r_1, \ldots, r_{k-2}\}$ where $u \in S_{t-1}$ and $u' \in A\setminus (S_{t-1}\cup \oS_{t-1})$,
 and let $w$ be the node it corresponds to in $H_t$.
 For each vertex $y' \in A\setminus (S \cup \oS_{t-1})$ such that
 $(u,y')\notin \ovE_{t-1}$ and
  $(y',r_p) \notin \ovE_{t-1}$ for every  $p\in [k-2]$, there is a neighbor $\bar{w}$ of $w$ in $H_t$, corresponding to the graph $\hG'\in \ncalW_t$ that results from $\hG$ by performing the following operations.

Let $v_{1},\dots,v_{k-1}$ be the neighbors of $y'$ in $C$.
First, we remove the edge $(u,u')$ and add the edge $(y,y')$.
Next,  we remove the edge $(u',r_p)$ for each
$p\in[k-2]$, and the edge $(y',v_{p})$ for each $p\in [k-1]$.
Finally,  we
add the edges $(y',r_p)$ for each $p \in [k-2]$ and the edges $(u',v_{p})$ for each $p \in [k-1]$.

It remains to specify the labels of all new  edges.
The  new (oriented) edge $(u,y')$ is given the label of the removed edge $(u,u')$, and $(y',u)$ is given the label of $(y',v_{k-1})$.
Each new edge $(y',r_p)$ for $p\in [k-2]$ is given the label of the removed edge $(y',v_{p})$,
each new edge $(u',v_{p})$ for $p \in [k-2]$ is given the label of the removed edge $(u',r_p)$, and
$(u',v_{k-1})$ is given the label of $(u',u)$.

 	As before, by the description of the process $\cP$, it holds that
 	\[
 	\Pr[\calE^1_t\;|\; q_t = \pair(u,u')]=\frac{|\calW_t)|}{|\cG_t|}=\frac{|W|}{|W\cup \oW|}.
 	\]
 	Let $d_w, d_{\ow}$ denote the average degrees of the nodes in $W$ and $\oW$, respectively.
 	Since $H_t$ is a bipartite graph, $|W| =\frac{|\oW|\cdot d_{\ow}}{d_w}$. Hence, to bound $\Pr[a_t=\mW_2]$, we will be interested in bounding $|W|$ by lower bounding $d_w$ and upper bounding $d_{\ow}$.

 By the definition of the set of neighbors of each node $w$ in $W$,
the number of ``eligible'' vertices $y'$  that can be used to replace $u'$ in the hidden clique and hence define a neighbor $\bar{w}$ of $w$
is lower bounded by $|A| - k - |\oS_{t-1}| - |\ovE_{t-1}|$.
Since 
$|\oS_{t-1}| + |\ovE_{t-1}| \leq \ell/\cnsta$,
there are at least $\ell/\cnstb$ such vertices.
Therefore, $d_W \geq \ell/\cnstb$. 

Now consider a non-witness graph $\hG' \in \ncalW_t$.
Since $\hG'$ is a non-witness graph, its special set  $S' = \{u,r_1, \ldots, r_{k-1}\}$ of $\hG'$ contains $u$ but not $u'$. By the definition of $H_t$, each neighbor in $H_t$ of the node $\bar{w}_{\hG'}$ corresponding to $\hG'$ must correspond to a graph $\hG \in \calW_t$ that contains a $k$-clique over $u,u'$ and $k-2$ of the vertices among $\{r_1,\dots,r_{k-1}$. Therefore, $\bar{w}_{\hG'}$ has at most $k-1$ neighbors in $H_t$. Since this is true for every node $\bar{w} \in \oW$, we have that $d_{\oW} < k$.

 	Therefore,  $|W| \leq \cnstb|\oW|\cdot k/\ell$, and so
 	\[
 	\Pr[\calE^1_t \;|\; q_t=\pair(u,u')]=\frac{|W|}{|W\cup \oW|}\leq \frac{|W|}{|\oW|}\leq \frac{\cnstb k}{\ell}\;.
 	\]
As noted above (and similarly to the proof of Claim~\ref{clm:witness-edge}) showing that
$\Pr[\calE^1_t \;|\; q_t =\nbr(u,i) \;\&\; a_t=u'] \leq \cnstb k/\ell$ is the same, except that in each witness graph, the edge $(u,u')$ must be labeled $i$.
 \end{proof}

 We are finally ready to prove our main lower bound theorem.
 \begin{proof}[Proof of Theorem~\ref{thm:lb}.]
 Recall that we assume for the sake of the analysis, that if the algorithm completes its execution after $T$ queries and $|S_T| < k$, then in order to select the final $\hG \in \cG_T$, the process first determines whether  the vertices in $\widehat{S}\setminus S_T$ belong to the hidden clique
 $S(\hG)$ of $\hG$ (in the same manner that it answers queries).
 	Recall that by the definition of the witness events, for any $t'$,
 unless the event $\calE^2_{t}$ has occurred for some $t<t'$, the event $\calE^1_{t'}$ cannot occur.


Therefore, for the algorithm to succeed, it must hold that for some $k_2\geq 1$ and $k_1 \geq 0$ such that $2k_2+k_1 = k$, there are $k_2$ indices $t_1,\dots,t_{k_2}$ and $k_1$ indices $t'_1,\dots,t'_{k_1}$, such that
the events $\calE^2_{t_j}$ for $j \in [k_2]$ all hold  and the events for $j\in [k_1]$ $\calE^1_{t'_j}$ all hold.
By Claim~\ref{clm:witness-edge} and Claim~\ref{clm:witness-vertex} (and since
$\frac{T\cdot k^2}{\ell^2}< \left(\frac{T\cdot k}{\ell}\right)^2$ for $x > 1$) 
 	\[
 	\Pr[\widehat{S}=S(\hG)]\leq
 \frac{\cnstb  T \cdot k^2}{\ell^2} \cdot \left(\frac{\cnstb T\cdot k}{\ell} \right)^{k-2}
  = \frac{(\cnstb T)^{k-1}\cdot k^k}{\ell^k}\;.
 	\]
 	By the setting of $\ell=\Theta(\sqrt{n\alpha})$, in order to have $\Pr[\widehat{S}=S(\hG)]=\Omega(1/n_k)$, it must hold that $T=\Omega\left( \left(\frac{(n\alpha)^{k/2}}{k^k \cdot n_k} \right)^{\frac{1}{k-1}}\right)$.
\end{proof}

\subsection{The counting-based lower bound}\label{subsec:lb-counting}
The second term in the lower bound of Theorem~\ref{thm:lb} follows directly from a lower bound of 
\begin{equation}
\label{eqn:ers20-lb}
\Omega\left(\min\left\{n\alpha, \frac{n(\alpha/k)^{k-1}}{n_k}\right\}\right)
\end{equation}
by~\cite{ERS20_soda} for $k$-clique counting. 
They prove that any algorithm that performs fewer queries than~(\ref{eqn:ers20-lb}) cannot distinguish between two families of graphs with high probability: one with $n_k$ $k$-clique, and one with no $k$-cliques. Since any $k$-clique sampling algorithm could distinguish between the two families (by returning a $k$-clique if a the graph belongs to the former family), any uniform sampling algorithm cannot perform fewer queries.

\ifnum\fullversion=0

\fi

\bibliographystyle{alpha}
\bibliography{all_bib_combined}

\newcommand{\etalchar}[1]{$^{#1}$}
\begin{thebibliography}{BHPR{\etalchar{+}}17}

\bibitem[ABG{\etalchar{+}}18]{Aliak}
Maryam Aliakbarpour, Amartya~Shankha Biswas, Themis Gouleakis, John Peebles,
  Ronitt Rubinfeld, and Anak Yodpinyanee.
\newblock Sublinear-time algorithms for counting star subgraphs via edge
  sampling.
\newblock {\em Algorithmica}, 80(2):668--697, 2018.

\bibitem[AKK18]{AKK19}
Sepehr Assadi, Michael Kapralov, and Sanjeev Khanna.
\newblock {A Simple Sublinear-Time Algorithm for Counting Arbitrary Subgraphs
  via Edge Sampling}.
\newblock In Avrim Blum, editor, {\em 10th Innovations in Theoretical Computer
  Science Conference (ITCS 2019)}, volume 124 of {\em Leibniz International
  Proceedings in Informatics (LIPIcs)}, pages 6:1--6:20, Dagstuhl, Germany,
  2018. Schloss Dagstuhl--Leibniz-Zentrum fuer Informatik.

\bibitem[BBGM19]{bhattacharya2019triangle}
Anup Bhattacharya, Arijit Bishnu, Arijit Ghosh, and Gopinath Mishra.
\newblock Triangle estimation using tripartite independent set queries.
\newblock In {\em 30th International Symposium on Algorithms and Computation
  (ISAAC 2019)}. Schloss Dagstuhl-Leibniz-Zentrum fuer Informatik, 2019.

\bibitem[BHPR{\etalchar{+}}17]{beame2017edge}
Paul Beame, Sariel Har-Peled, Sivaramakrishnan~Natarajan Ramamoorthy, Cyrus
  Rashtchian, and Makrand Sinha.
\newblock Edge estimation with independent set oracles.
\newblock {\em arXiv preprint arXiv:1711.07567}, 2017.

\bibitem[CLW19]{Chen-IS-edges}
Xi~Chen, Amit Levi, and Erik Waingarten.
\newblock Nearly optimal edge estimation with independent set queries.
\newblock {\em CoRR}, abs/1907.04381, 2019.

\bibitem[DLM20]{DLM20}
Holger Dell, John Lapinskas, and Kitty Meeks.
\newblock Approximately counting and sampling small witnesses using a colourful
  decision oracle.
\newblock In {\em Proceedings of the Thirty-First Annual ACM-SIAM Symposium on
  Discrete Algorithms}, pages 2201–--2180, 2020.

\bibitem[ELRS15]{ELRS-focs}
Talya Eden, Amit Levi, Dana Ron, and C~Seshadhri.
\newblock Approximately counting triangles in sublinear time.
\newblock In {\em Foundations of Computer Science , 2015 IEEE 56th Annual
  Symposium on}, pages 614--633. IEEE, 2015.

\bibitem[ER18a]{ER18_LB}
Talya Eden and Will Rosenbaum.
\newblock Lower bounds for approximating graph parameters via communication
  complexity.
\newblock In {\em Approximation, Randomization, and Combinatorial Optimization.
  Algorithms and Techniques 2018}, pages 11:1--11:18, 2018.

\bibitem[ER18b]{ER18_edges}
Talya Eden and Will Rosenbaum.
\newblock On sampling edges almost uniformly.
\newblock In {\em 1st Symposium on Simplicity in Algorithms, {SOSA} 2018,
  January 7-10, 2018, New Orleans, LA, {USA}}, pages 7:1--7:9, 2018.

\bibitem[ERR19]{ERR19}
Talya Eden, Dana Ron, and Will Rosenbaum.
\newblock The arboricity captures the complexity of sampling edges.
\newblock In {\em 46th International Colloquium on Automata, Languages, and
  Programming, {ICALP} 2019, July 9-12, 2019, Patras, Greece.}, pages
  52:1--52:14, 2019.

\bibitem[ERS19]{ERS19-sidma}
Talya Eden, Dana Ron, and C.~Seshadhri.
\newblock Sublinear time estimation of degree distribution moments: The
  arboricity connection.
\newblock {\em {SIAM} J. Discrete Math.}, 33(4):2267--2285, 2019.

\bibitem[ERS20a]{ERS20_soda}
Talya Eden, Dana Ron, and C.~Seshadhri.
\newblock Faster sublinear approximation of the number of \emph{k}-cliques in
  low-arboricity graphs.
\newblock In {\em Proceedings of the 2020 {ACM-SIAM} Symposium on Discrete
  Algorithms, {SODA} 2020, Salt Lake City, UT, USA, January 5-8, 2020}, pages
  1467--1478, 2020.

\bibitem[ERS20b]{ERS20-clqs}
Talya Eden, Dana Ron, and C.~Seshadhri.
\newblock On approximating the number of $k$-cliques in sublinear time.
\newblock {\em SIAM Journal on Computing}, 49(4):747--771, 2020.

\bibitem[Fei06]{feige2006sums}
Uriel Feige.
\newblock On sums of independent random variables with unbounded variance and
  estimating the average degree in a graph.
\newblock {\em SIAM Journal on Computing}, 35(4):964--984, 2006.

\bibitem[FGP20]{Peng}
Hendrik Fichtenberger, Mingze Gao, and Pan Peng.
\newblock Sampling arbitrary subgraphs exactly uniformly in sublinear time.
\newblock In {\em 47th International Colloquium on Automata, Languages, and
  Programming, {ICALP} 2020, July 8-11, 2020, Saarbr{\"{u}}cken, Germany
  (Virtual Conference)}, pages 45:1--45:13, 2020.

\bibitem[GR08]{GR08}
Oded Goldreich and Dana Ron.
\newblock Approximating average parameters of graphs.
\newblock {\em Random Structures \& Algorithms}, 32(4):473--493, 2008.

\bibitem[GRS11]{GRS11}
Mira Gonen, Dana Ron, and Yuval Shavitt.
\newblock Counting stars and other small subgraphs in sublinear-time.
\newblock {\em SIAM Journal on Discrete Mathematics}, 25(3):1365--1411, 2011.

\bibitem[JVV86]{jerrum1986random}
Mark~R Jerrum, Leslie~G Valiant, and Vijay~V Vazirani.
\newblock Random generation of combinatorial structures from a uniform
  distribution.
\newblock {\em Theoretical computer science}, 43:169--188, 1986.

\bibitem[NW64]{nash1964decomposition}
C.~St.~JA. Nash-Williams.
\newblock Decomposition of finite graphs into forests.
\newblock {\em Journal of the London Mathematical Society}, 1(1):12--12, 1964.

\bibitem[Sch81]{schnorr1981self-transformable}
C.~P. Schnorr.
\newblock {\em On self-transformable combinatorial problems}, page 225–243.
\newblock Mathematical Programming Studies. Springer, 1981.

\bibitem[T{\v{e}}t20]{Tetek}
Jakub T{\v{e}}tek.
\newblock Sampling an edge uniformly in sublinear time.
\newblock {\em arXiv preprint arXiv:2009.11178}, 2020.

\end{thebibliography}

\appendix
\section{Missing proofs from Section~\ref{sec:ub}}\label{sec:appendix}

\begin{proof}[Proof of Claim~\ref{clm:Hk_arb}.]
	By~\cite{nash1964decomposition},
	for any graph $F$, $\alpha(F)=\max_{F'\subseteq F}\left\lceil
	\frac{m_{F'}}{n_{F'}-1}
	\right\rceil,$ where $n_{F'}$ and $m_{F'}$ denote the number of vertices and edges in the subgraph $F'$.
	Assume towards contradicition that $H_k$ has arboricity $\alpha'>\alpha$, and let $H'$ be a  subgraph of $H_k$ such $\alpha(H_k)=\left\lceil
	\frac{m_{H'}}{n_{H'}-1}
	\right\rceil$.
	Let $A$ be the set of all vertices of $G$ that participate in the nodes of $H'$.
	Then the subgraph $G[A]$ has $n_{k-1}(G[A])= n_{H'}$ $(k-1)$-cliques, and \[2n_{k}(G[A])\geq 2m_{H'}>\alpha' \cdot n_{H'}>\alpha \cdot n_{k-1}(G[A])\]
	$k$-cliques. But this is a contradiction to Claim~\ref{clm:bound-clqs}. Hence the claim follows.
\end{proof}

\medskip
In order to prove Lemma~\ref{lem:rw}
we first describe a decomposition of a graph's vertices into 
disjoint \emph{layers} $L_0,\dots,L_s$.
Layers $L_1,\dots,L_s$ are determined given
the zeroth layer $L_0$ and a parameter $\beta \in (0,1)$.
This decomposition 
is essentially the same as the decomposition of~\cite{ERR19}, except that here we do not predetermine $L_0$.

\begin{dfn}[Layering of a graph]\label{def:layering_low}
	Let $F=(V_F,E_F)$ be a graph, let $L_0 \subseteq V_F$ be a subset of vertices, and let
	$\beta\in(0,1)$ be a parameter.
	Starting from the given $L_0$, we define an $(L_0,\beta)$-layering of  the vertices of
	$F$ into a series of non-empty disjoint layers $L_0, L_1, \ldots, L_s$, defined iteratively  as follows.
	$L_0$ is as given, and for $j\geq 1:$		
	\begin{align}
	L_j=\{ v\in V_F \;:\; |\Gamma_F(v)\cap (L_0\cup \ldots \cup L_{j-1})| >(1-\beta)|\Gamma_F(v)|\; \} \;. \label{eq:Lj}
	\end{align}
	That is, $L_j$ is the set of vertices for which $(1-\beta)$ of their neighbors reside in $L_0\cup \ldots \cup L_{j-1}$.
	We say that a graph $F=(V_F,E_F)$ admits   an $(L_0,\beta)$-layering of \emphdef{depth} $s$  if $V_F=L_0\cup \ldots \cup L_{s}$.
\end{dfn}

\begin{ntn}
	For a graph $F$ and a decomposition as above,  let $E_F(L_j)$ denote the set of oriented edges incident to the vertices of $L_j$, $E_F(L_j)=\{(v,u)\mid v\in L_j \}$. We sometimes omit the subscript $F$ when the graph at question is clear from context.
\end{ntn}

We claim that for any choice of parameter $\beta$ and an appropriate choice of the layer $L_0$, the decomposition in the above definition has depth $s=\lceil \log|V_F|\rceil$.

\begin{clm}\label{clm:depth}
	Let $F$ be a graph of arboricity $\alpha$, and let $\beta \in (0,1)$ be a parameter.
	Suppose that for $\tau$ that satisfies $\tau\cdot \beta\geq 4\alpha$, we have that
	for every $v$ such that $d_F(v) \leq \tau$ it holds that $v\in L_0$.
	Then the graph $F$ admits an $(L_0,\beta)$-layering of depth $s=\lceil \log |V_F| \rceil$
	as defined in Definition~\ref{def:layering_low}.
\end{clm}
\begin{proof} 
	For each $i$, let $W_i = V_F \setminus (L_0 \cup L_1 \cup \cdots \cup L_{i-1})$ be the set of vertices not in levels $0, 1, \ldots, i-1$. Let $m(W_i)$ denote the number of edges in the subgraph of $F$ induced by $W_i$.
	For any fixed $i$ and $v \in W_{i+1}$, we have $d_{< i}(v) < (1 - \beta) d(v)$ because $v \notin L_{\leq i}$. Therefore, $v$ has at least $\beta d(v) > \beta \tau$ neighbors in $W_i$ (recall that if $v\notin L_0$ then by the promise that $L_0$ contains all vertices with degree at most $\tau$, $d(v)\geq \tau$). Summing over vertices $v \in W_{i+1}$ gives
	\begin{equation}
	m(W_i) = \frac{1}{2}\sum_{v\in W_i}d_{\geq i}(v) \geq \frac{1}{2}\sum_{v\in W_{i+1}}d_{\geq i}(v) > \frac{1}{2} \abs{W_{i+1}}\cdot \beta\tau\;.
	\label{eq:mWi-lb}
	\end{equation}
	On the other hand, since $G$ has arboricity at most $\alpha$, Theorem~\ref{clm:bound-clqs} implies that
	$m(W_i) \leq \alpha \abs{W_i}\;.$
	Combining the above upper bound with Equation~(\ref{eq:mWi-lb}), it follows that
	$\frac{\abs{W_{i+1}}}{\abs{W_i}} \leq \frac{2 \alpha}{\beta \tau} \leq  \frac 1 2\;,$
	where the last inequality is by the assumption that  of $\beta\cdot \tau\geq4\alpha$. Therefore, $s \leq \lceil \log |V_F| \rceil$, as required.
\end{proof}

We shall use the following notation in the proof of Lemma~\ref{lem:rw}.
\begin{ntn} \label{def:P}
	For a graph $F=(V_F, E_F)$ and an  edge $e\in E_F$, let  $\HP_j[e]$ denote the probability that \se\ returns $e$ when the chosen index in Step~\ref{step:se_j} is $j$.
	Further let $\HP_{\leq j}[e] \eqdef \sum_{i=0}^{j} \HP_i[e]$  and similarly for $\HP_{\geq j}[e]$.	
\end{ntn}

\begin{proof}[Proof of Lemma~\ref{lem:rw}]\label{sec:proof_rw}
	By the assumption that \ISLZ\ is a $(\delta, \tau)$-good $L_0$-oracle for $\delta\leq \beta'/X$, it holds that with probability at least $1-\delta\geq1-\beta'/X$, \ISLZ\ determines a $\tau$-good $L_0$, (see Definition~\ref{def:good_L0}). Denote this event by $\mathcal{E}$, and condition on it holding.
	
	To prove the claim we shall prove the following two items.
	\begin{itemize}
		\item For every $j \in [s]$, $e\in E_F(L_j)$ and $\ell \in [j, s]$, it holds that  $\HP_{\leq \ell}[e] \geq \frac{(1-\beta)^{2j+1}}{X}$. 
		\item For every $e\in E_F$, $\HP_{\leq s}[e]\leq \frac{(1+\beta)^{j+1}}{X}$.
	\end{itemize}
	
	We start with the first item which we prove by induction on $j$, starting with $j=0$.
	Let $e=(u,u')$ be an edge in $E_F(L_0)$  (so that $u\in L_0(F)$).
	By the assumption that \slz\ is a good $L_0$-sampling subroutine for $F$, it holds that the invocation 
\slz$( F, \oN_F,\beta',\tau,\vec{p})$
in Step~\ref{step:sallow} returns $e$ with probability  $\frac{1\pm \beta'}{X}$.
	Hence,
	\begin{equation}
	\HP_0[e]\geq\frac{(1-\beta')}{X} \;. \label{eqn:P0}
	\end{equation}
	Hence, for $j=0$ and $0<\ell \leq s$,
	\begin{equation}
	\HP_{\leq \ell}[e]=\sum_{i=0}^{\ell} \HP_{i}[e] = \HP_0[e]+ \sum_{i=1}^{\ell}\HP_{i}[e]\geq  \frac{(1-\beta')}{X}\;. \label{eqn:P}
	\end{equation}
	We now assume that the first item in the claim holds for all $i \leq j-1$ and $\ell \in [i, s]$, and prove that it holds for $j$ and for every $\ell \in [j, s]$.
	By the induction hypothesis, for every $i \leq j-1$, $(v,u)\in E_{F}(V_i)$, $P_{\leq j-1}[v,u]\geq \frac{(1-\beta')^{2i+2}}{X}$.
	Moreover,  by the assumption that \snbr\ is a $\beta'$-good neighbor-sampling subroutine with respect to $F$, for every $u'\in \Gamma_F(u)$, the probability that $u'$ is returned when \snbr\ is invoked with $u$ is at least $\frac{1-\beta'}{d(u)}$.
	Therefore, for every $e=(u,u') \in E(V_j)$,
	\begin{align*}
	\HP_{\leq \ell}[e]&\geq \HP_{\leq j}[e]
	\;=\;\sum_{(v,u)  \in E_F}\HP_{\leq j-1}[(v,u)]\cdot \Pr[\snbr\ \text{returns $u'$}]
	\\&\geq \sum_{(v,u)  \in E_F}\HP_{\leq j-1}[(v,u)]\cdot \frac{1-\beta'}{d(u)}
	\;\geq\; \sum_{i=0}^{j-1}\sum_{(v,u)\in E_F(V_i)}\HP_{\leq j-1}[(v,u)]\cdot \frac{1-\beta'}{d(u)}\\
	&\;\geq \; \sum_{i=0}^{j-1}\sum_{(v,u)\in E_F(V_i)} \frac{(1-\beta')^{2i+1}}{X}\cdot \frac{1-\beta'}{d(u)}
	\geq\;\frac{(1-\beta')^{2j-1} \cdot d_{\leq j-1}(u)}{X}\cdot \frac{1-\beta'}{d(u)}
	\\& \geq \frac{(1-\beta')^{2j+1}}{X},
	\end{align*}
	where we used the decomposition property of Definition~\ref{def:layering_low}, that $d_{\leq j-1}(u) > (1-\beta')d(u)$.
	Hence, the first item of the claim holds for every $j \in [s]$ for every $\ell \in [j, s]$.
	
	We continue to prove the second item. We prove by induction on $j$ that for every $j\in [s]$, $\HP_{\leq j}[e]\leq \frac{(1+\beta')^{j+1}}{X}$.
	First for $e\in E_F(L_0)$, by the assumption that \slz\ is a good $L_0$-sampling subroutine,
	it holds that
	\(
	\HP_0[e]\leq \frac{1+\beta'}{X}.
	\)
	Furthermore, by the condition on $\mathcal{E}$, \slz\ is a $\tau$-good $L_0$-sampling subroutine, only edges in $E_F(L_0)$ are returned in Step~\ref{step:sallow}. Hence,   for every $e \notin E_F(L_0)$,
	\(
	\HP_j[e]=0.
	\)
	Therefore, the claim holds for every $e$ and $j=0$.
	We now assume that for every $e$, for every $i\leq j-1$, $\HP_{\leq i}[e] \leq \frac{(1+\beta')^{i+1}}{X}$, and prove it for $j$.
	First, for $e\in L_0$, due to Step~\ref{step:check_L0},
	$e$ can only be returned if \se\ is invoked with $j=0$, in which case, $e$ is returned with probability at most $\frac{1+\beta'}{X}$.
	Hence,
	\[
	\HP_{\leq j}[e]=\sum_{i=0}^s\HP_j[e]= \HP_0[e]\leq \frac{1+\beta'}{X}.
	\]
	Now consider an edge $e=(u,u')\notin E_F(L_0)$.
	We deal separately with $\HP_0[e]$ and $\HP_{\leq j}[e]-\HP_0[e]$.	
	By the assumption that \slz\ is a $(\beta',\tau)$-good $L_0$-sampler,  $\HP_0[e]=\frac{\beta'}{X}$.
	Also, by the assumption that \snbr\ is a $\beta'$-good neighbor-sampling subroutine, if \snbr\ is invoked with a node $u\in V_F$, then for every $u' \in \Gamma_F(u)$, the probability that it is returned by the subroutine is at most $\frac{1+\beta'}{d(u)}$.
	Hence,
	\begin{align*}
	\HP_{\leq j}[e]&=\frac{\beta'}{X}+\sum_{i=1}^j\HP_i[e]\leq \sum_{i=1}^j \sum_{(v,u)\in E_F} \HP_{i-1}[(v,u)]\cdot \frac{1+\beta'}{d(u)} =\sum_{(v,u) \in E_F} \frac{1+\beta'}{d(u)} \sum_{i=0}^{j-1}\HP_{i}[(v,u)]\\
	& =\sum_{(v,u) \in E_F} \frac{1+\beta'}{d(u)} \cdot \HP_{\leq j-1}[(v,u)]\leq \frac{(1+\beta')^{j}}{d(u)}\cdot \frac{(1+\beta')}{X}\cdot d(u)=\frac{(1+\beta')^{j+1}}{X}.
	\end{align*}
	Therefore, $\HP_{\leq j}[e]\leq \frac{(1+\beta')^{j+1}}{X}$.
	This concludes the proof of the second item.
	
	For every edge $e\in F$, let $\HP_{i}[e\mid \mathcal{E}]$  ($\HP_i[e\mid \overline{\mathcal{E}}]$) denote the events that the edge $e$ is returned in the $i\th$  invocation, conditioned on the event $\mathcal{E}$ ($\overline{\mathcal{E}}$).
	Then 
	\[
	\HP_{\leq s}[e]=\HP_{\leq s}\Pr[e\mid \mathcal{E}]\cdot \Pr[\mathcal{E}]+ \HP_{\leq s}[e\mid \overline{\mathcal{E}}]\cdot \Pr[\overline{\mathcal{E}}]\;.
	\]
	Hence, for $\delta=\beta'/X$, 	$\HP_{\leq s}[e] \geq \left(1-\delta\right)\frac{(1-\beta')^{2j+1}}{X}\geq \frac{(1-\beta')^{2j+2}}{X}$
	and $\HP_{\leq s}[e] \leq \frac{(1+\beta')^{j+1}}{X}+\delta\leq \frac{(1+\beta')^{j+2}}{X}.$
	Therefore,for every edge $e\in F$, $P_{\leq s}[e]\in \left[\frac{(1-\beta')^{2s+2}}{X}, \frac{ (1+\beta')^{s+2}}{X}\right]$.
	Consider a specific edge $e \in F$, and assume that $v \in L_i$ for some $i\in [0, \ldots, s]$, $s=\left\lceil\log \overline{N}_F\right\rceil$. By the above,
	\begin{align*}
	\Pr[e \text{ is returned}]&=\sum_{i=0}^{s} \Pr[j=i]\cdot 	\Pr[e \text{ is returned}\mid  j=i] \\
	&= \sum_{i=0}^{s} \frac{1}{s+1}	\cdot P_{i}[e]=\frac{P_{\leq s}[e]}{s+1} \in  \left[(1-\beta')^{2s+1}, 1+\beta'\right]\cdot \frac{1}{X \cdot(s+1)}.
	\end{align*}
	Therefore, by the setting of $\beta'=\frac{\beta}{(2s+2)}$, for every edge $e$, 	$\Pr[e \text{ is returned}]\in \frac{1\pm\beta}{X\cdot (s+1)}$,
	as claimed.
\end{proof}

\medskip
\begin{proof}[Proof of Claim~\ref{clm:taylor}]\label{proof:taylor}
	By the Taylor expansion of the function $(1-x)^{\lceil y \rceil}$, it holds that
	\[
	1-x\lceil y\rceil \leq (1-x)^{\lceil y \rceil} \leq 1-x\lceil y\rceil +(x\lceil y\rceil)^2 .
	\]
	Therefore,
	\[
	1-(1-x)^{\lceil y \rceil} \leq x\lceil y\rceil \leq x(y+1)=xy(1+1/y).
	\]
	And
	\[	1-(1-x)^{\lceil y \rceil}	\geq x\lceil y\rceil -x^2(\lceil y \rceil)(\lceil y \rceil-1)\geq xy-2x^2y^2 =xy(1-2xy).\]
	
\end{proof}

\end{document}